\documentclass[a4paper,prd,twocolumn,preprintnumbers,superscriptaddress,nofootinbib,amsmath,amssymb]{revtex4-2}
\usepackage[dvips]{graphics}
\usepackage{color,hyperref}
\usepackage{times}
\usepackage{mathrsfs}
\hypersetup{colorlinks=true,linkcolor=blue,citecolor=cyan,filecolor=magenta,urlcolor=blue}
\usepackage{enumitem}
\usepackage{graphicx,subfigure}
\usepackage{dcolumn}
\usepackage{bm}
\usepackage{color}
\usepackage{babel}
\usepackage{orcidlink}     

\begin{document}

\title{Spinning particle dynamics around a non-singular Schwarzschild black hole in conformal gravity}

\author{Shavkat Karshiboev}\email{shavkatarshiboyev@samdpi.uzl}
\affiliation{Samarqand State Pedagogical Institute, Spitamen Shokh Street 166, Samarkand 140100, Uzbekistan}

\author{Bobur Turimov}\email{bturimov@astrin.uz}
\affiliation{School of Physics, Harbin Institute of Technology, Harbin 150001, People’s Republic of China}
\affiliation{Engineering school, Central Asian University, Milliy Bog Str. 264, Tashkent 111221, Uzbekistan}
\affiliation{Ulugh Beg Astronomical Institute, Astronomy Str. 33, Tashkent 100052, Uzbekistan}

\author{Sulton Usanov}\email{sm.usanov@kiut.uz} 
\affiliation{Kimyo International University in Tashkent, Shota Rustaveli street 156, Tashkent 100121, Uzbekistan}

\author{Ahmadjon Abdujabbarov}
\email{ahmadjon@astrin.uz}
\affiliation{School of Physics, Harbin Institute of Technology, Harbin 150001, People’s Republic of China}
\affiliation{Tashkent State Technical University, Tashkent 100095, Uzbekistan}

\author{Chengxun Yuan}
\email{yuancx@hit.edu.cn}
\affiliation{School of Physics, Harbin Institute of Technology, Harbin 150001, People’s Republic of China}

\author{Hasanova Sumanbar}\email{xasanovasumanbar@gmail.com}\affiliation{Department of Applied Sciences in Primary Education, Chirchik State Pedagogical University, 104 A. Temur Str., Chirchik 111700, Uzbekistan}

\author{Gulhayo Hayitova}\email{gulhayohayitova552@gmail.com}\affiliation{Institute of Nuclear technology, Samarkand State University named after Sh. Rashidov, Samarkand, 140104, Uzbekistan}

\date{\today}
\begin{abstract}
Spinning test particles provide a precision probe of strong-field gravity, yet their dynamics around singularity-free black holes in conformal gravity remain poorly constrained. We investigate equatorial motion around the non-singular Schwarzschild black hole of conformal gravity, in which the conformal scale \(L\) removes the curvature divergence at the origin while preserving the event horizon at \(r=2M\) and recovering the Schwarzschild geometry as \(L\to 0\). Using the Mathisson-Papapetrou-Dixon equations with the Tulczyjew spin supplementary condition, we derive the conserved energy and angular momentum in the conformally rescaled spherical spacetime and construct the radial effective potential for spinning particles. This allows us to determine the physical timelike spin bound, the innermost stable circular orbit (ISCO), the degeneracy between \(L\) and the Kerr rotation parameter, and representative bound trajectories. We show that \(L\) strengthens the effective-potential barrier, decreases the ISCO radius, angular momentum and energy, and lowers the maximum admissible spin before the trajectory becomes spacelike. The orbital integrations reveal a clear division of roles: \(L\) governs the compactness of the bound orbit, whereas the particle spin governs the accumulated relativistic precession. These results establish spinning-particle dynamics as a sensitive diagnostic of conformally regular black-hole spacetimes and provide a possible route to distinguish the conformal Schwarzschild geometry from standard Schwarzschild background.
\end{abstract}


\maketitle

\section{Introduction}

General Relativity (GR) remains the most successful classical theory of gravity and has passed stringent tests from Solar-System dynamics to gravitational-wave observations and horizon-scale imaging of compact objects \cite{Will2014,Berti2015,Abbott2016,EHT2019,EHT2022}. Nevertheless, classical GR predicts geodesic incompleteness and curvature singularities under broad physical assumptions \cite{Penrose1965,HawkingPenrose1970}. This limitation motivates the study of alternative gravitational frameworks and effective black-hole geometries in which strong-field observables can be compared with the Schwarzschild and Kerr predictions.

Conformal gravity provides a particularly interesting route to the singularity problem because the Weyl symmetry permits metrics related by local conformal transformations to represent the same gravitational configuration in the unbroken phase \cite{MannheimKazanas1989,Dabrowski2009,Clifton2012,ConformalThermo2022,Cai2024,Zhang2023Conformal,Bambi2019QNM,Wei2024}. Bambi, Modesto and Rachwal showed that a suitable conformal rescaling of the Schwarzschild and Kerr metrics can generate exact black-hole solutions that are free of curvature singularities and geodesically complete \cite{BambiModestoRachwal2017,Bambi:2017PRD}. For the spherically symmetric case, the geometry is conformally equivalent to Schwarzschild and contains a length scale $L$ through the conformal factor $S(r)=(1+L^2/r^2)^2$. The Schwarzschild horizon remains at $r=2M$, while the former singular point $r=0$ is converted into a regular boundary that cannot be reached by massive or massless probes in a finite proper time or affine parameter \cite{BambiModestoRachwal2017,Zhang:2018EPJC,Bambi:2016JCAP}. This conformal Schwarzschild spacetime is therefore not a regular black hole generated by an additional matter source, such as nonlinear electrodynamics. Instead, the removal of the singularity follows from the conformal structure of the gravitational theory. The parameter $L$ controls the strength of the conformal deformation and recovers the standard Schwarzschild metric in the limit $L\rightarrow0$. Since conformal symmetry is expected to be broken in the low-energy phase, the choice of conformal factor can leave physical imprints on particle motion, circular orbits and strong-field observables \cite{BambiModestoRachwal2017,Bambi:2016JCAPa,AaToshmatov2017,Turimov:2018PRD}. 

A broader perspective on singularity resolution is provided by regular black-hole models in which the central singularity is replaced by a de Sitter core or by matter sources such as nonlinear electrodynamics. Bardeen’s original proposal and later Hayward, Dymnikova, and Ayón-Beato–García solutions illustrate that nonsingular cores can arise from different physical mechanisms, while more recent constructions such as black-bounce spacetimes and effective frameworks for regular black holes emphasize the model-dependent nature of the resulting geometry \cite{Bambi2023Book,CarballoRubio2023, Bardeen1968,Hayward2006,Dymnikova1992,AyonBeato1998,Simpson2019,CarballoRubio2018}. Unlike conformal gravity, where the Schwarzschild singularity is removed by a Weyl rescaling without introducing additional matter fields, these regular black holes typically modify the field equations or the matter sector. Comparing their observational signatures — shadows, quasinormal modes, lensing, and thermodynamic behavior — with those of conformally regular black holes is therefore important for identifying whether a given deviation from Schwarzschild is due to conformal symmetry or to a regular matter source.

Spinning test particles offer a complementary probe of regular black-hole geometries. Recent studies of spinning-particle motion around rotating regular black holes have shown that spin-curvature coupling can shift the ISCO, alter periastron precession, and change the timelike bound on the particle spin, sometimes mimicking the effects of black-hole rotation or of a regular core \cite{BenavidesGallego2019}. These effects are relevant for extreme-mass-ratio inspirals, where the small body’s spin can accumulate over many cycles and leave an imprint on gravitational-wave phasing, and for electromagnetic observations of Sgr A* and M87*, where the shadow and photon ring depend on the near-horizon geometry \cite{Cardoso2019, Barack2019, EHT2024Horndeski}. Incorporating spinning-particle dynamics into tests of conformal gravity therefore provides a way to break degeneracies among conformal scale, black-hole spin, and regular-core parameters that cannot be resolved by geodesic motion alone.

Test-particle dynamics offers a direct way to probe these imprints because the effective potential, circular-orbit energy, angular momentum and ISCO are determined by the metric functions. The geodesic approximation, however, neglects the intrinsic angular momentum of the orbiting body. For compact objects moving in the field of a much heavier black hole, finite-size effects already appear at pole--dipole order through the coupling between the particle spin and spacetime curvature. The covariant description is given by the Mathisson--Papapetrou--Dixon (MPD) equations \cite{Mathisson1937,Papapetrou1951,Dixon1970}, supplemented here by the Tulczyjew spin supplementary condition \cite{Tulczyjew1959}. In this framework, spin--curvature coupling modifies the effective potential, shifts the ISCO, changes orbital precession and imposes a physical timelike bound on the admissible spin \cite{Saijo1998,Hartl2003,Toshmatov2019,WitzanyPiovano2024}.

Spinning-particle motion has been studied in Schwarzschild, Kerr and a variety of modified or regular compact-object backgrounds \cite{Jefremov2015,Harms2016,AaAbdulxamidov2022,Umarov2025,Turakhonov2026,Jefremov2022,BenavidesGallego2022,Drummond2023}. These studies show that spin effects can be comparable, at the level of orbital diagnostics, to changes in the background geometry itself. The conformal Schwarzschild black hole is therefore an appropriate setting in which to ask whether the conformal scale $L$ can be separated from the intrinsic spin $s$ of the test body, and whether the ISCO shift caused by $L$ can mimic the effect of Kerr rotation.

In this work we study equatorial spinning-particle dynamics around the non-singular Schwarzschild black hole in conformal gravity. We construct the MPD conserved quantities for the conformally rescaled spherical metric, analyze the effective potential, determine the physical spin bound, compute the ISCO radius, angular momentum and energy, quantify the degeneracy between the conformal scale and the Kerr rotation parameter, and integrate representative trajectories. Throughout the paper we use geometrized units, $G=c=1$, and measure lengths in units of the black-hole mass $M$ unless stated otherwise.

\section{Spinning particle dynamics}
\label{section2}

\subsection{Non-singular spherical symmetric black hole}
\label{sec:nonsingular}

The background considered in this manuscript is the non-singular Schwarzschild black hole obtained in conformal gravity by a Weyl rescaling of the Schwarzschild metric \cite{BambiModestoRachwal2017}. In the static coordinates used throughout this work the line element is
\begin{equation}
ds^2=S(r)\left[-f(r)dt^2+\frac{dr^2}{f(r)}+r^2d\Omega^2\right],
\label{metric}
\end{equation}
where
\begin{equation}
f(r)=1-\frac{2M}{r},
\qquad
S(r)=\left(1+\frac{L^2}{r^2}\right)^2 .
\label{metric_function}
\end{equation}
Here $M$ is the black-hole mass and $L$ is the conformal length scale parameter. The Schwarzschild spacetime is recovered for $L=0$, while $L\neq0$ introduces a conformal deformation that is negligible at spatial infinity, since $S(r)\rightarrow1$ as $r\rightarrow\infty$.

The horizon is still located at $r_h=2M$ because the conformal factor is finite and positive there. By contrast, the former Schwarzschild singularity at $r=0$ is mapped to the point at which $S^{-1}(r)$ vanishes. The regularity of the geometry can be seen from the Ricci and the Kretschmann invariant~\cite{BambiModestoRachwal2017}:
\begin{align}
R&=-\frac{12L^2r(r-4 M)}{\left(L^2+r^2\right)^3}\ ,
\\\nonumber
K&=\frac{16r^2}{\left(L^2+r^2\right)^8}\Big[L^8\left(39 M^2-20 M r+3 r^2\right)\\\nonumber&+2 L^6 r^2 \left(42 M^2-16 M r+r^2\right)\\\nonumber&+L^4 r^4 \left(150 M^2-108 M r+23 r^2\right)\\&+12 L^2 M^2 r^6+3 M^2 r^8\Big]\ ,
\end{align}
which is finite everywhere and tends to zero at $r=0$. Similarly, one can show that the components of the Einstein tensor 
\begin{align}
&G^t_t=-\frac{4 L^2 r \left(L^2 M+3 M r^2-r^3\right)}{\left(L^2+r^2\right)^4}\ ,
\\
&G^r_r=\frac{4 L^2 r \left(-3 L^2 M+L^2 r+3 M r^2-2 r^3\right)}{\left(L^2+r^2\right)^4}\ ,
\\
&G^\theta_\theta=\frac{4 L^2 r \left(-4 L^2 M+L^2 r-6 M r^2+2 r^3\right)}{\left(L^2+r^2\right)^4}\ ,
\\
&G^\phi_\phi=\frac{4 L^2 r \left(-4 L^2 M+L^2 r-6 M r^2+2 r^3\right)}{\left(L^2+r^2\right)^4}\ .    
\end{align}
are regular at origin. The same spacetime is geodesically complete: massive particles, conformally coupled massive particles and photons do not reach $r=0$ in a finite proper time or affine parameter \cite{BambiModestoRachwal2017}. Hence the parameter $L$ should be interpreted as the scale controlling the conformal regularization of the Schwarzschild geometry rather than as a matter charge.
\begin{figure*}
\centering
\includegraphics[width=0.32\textwidth]{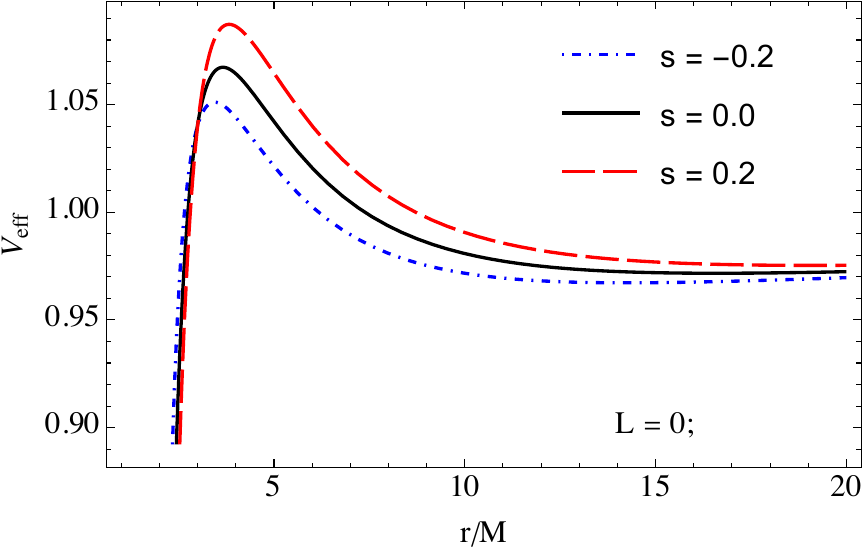}\hfill
\includegraphics[width=0.32\textwidth]{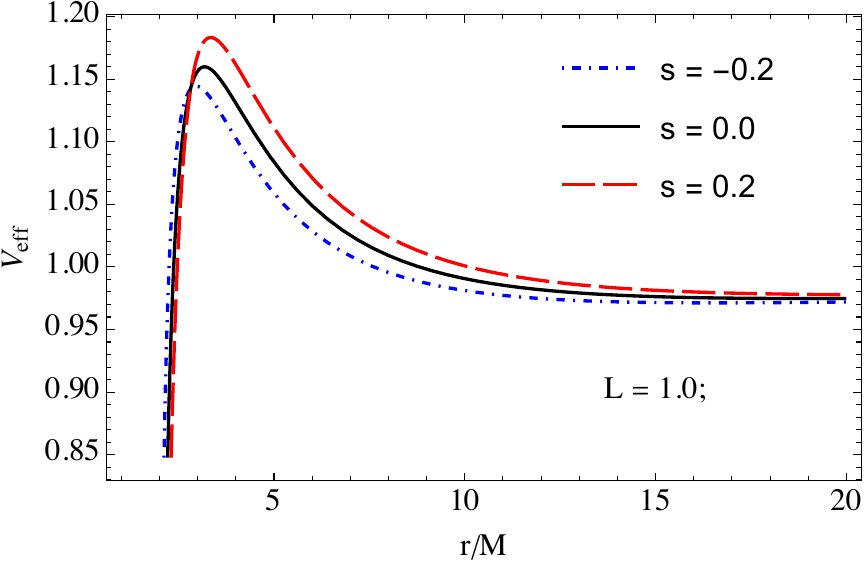}\hfill
\includegraphics[width=0.32\textwidth]{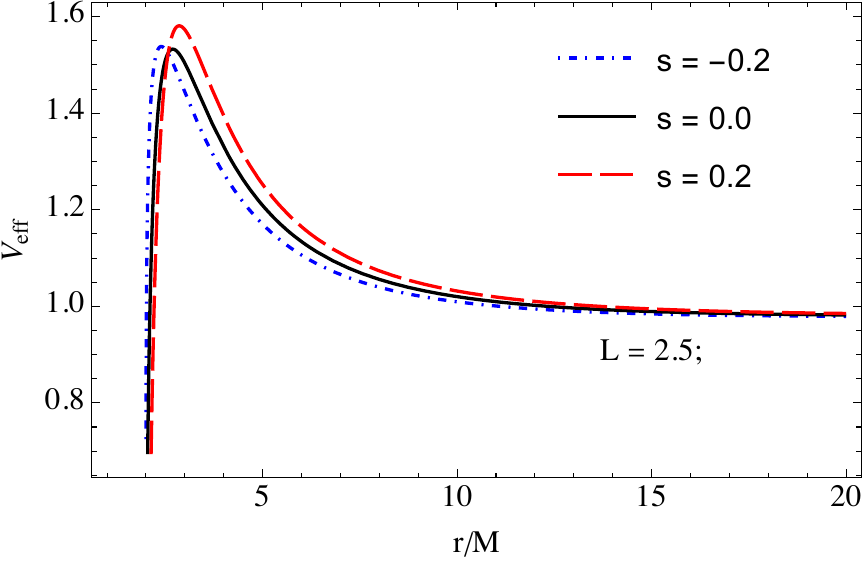}\\[0.5em]
\includegraphics[width=0.32\textwidth]{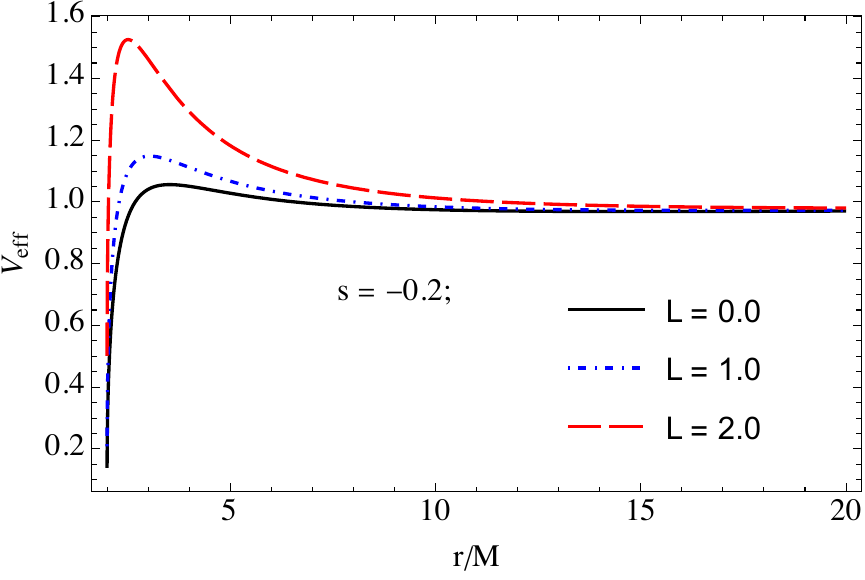}\hfill
\includegraphics[width=0.32\textwidth]{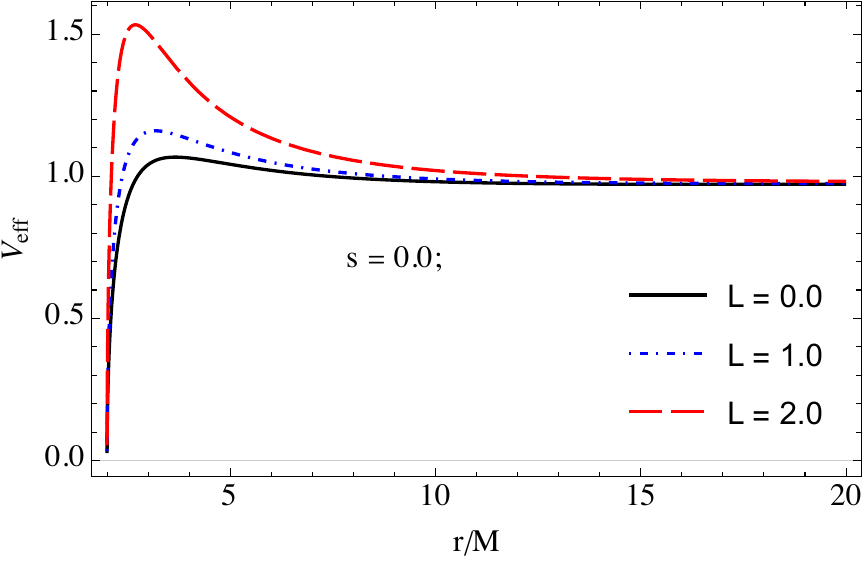}\hfill
\includegraphics[width=0.32\textwidth]{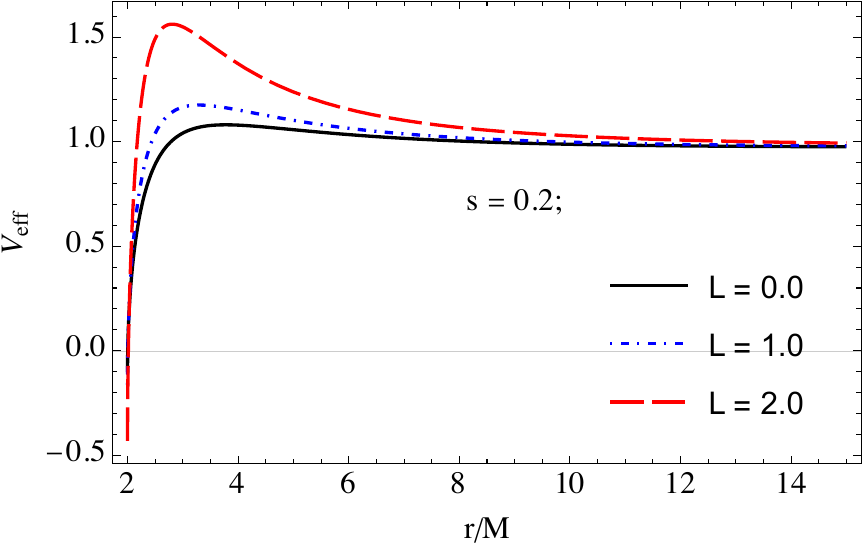}
\caption{Effective potential of spinning particles around the conformal Schwarzschild black hole. The upper row shows the dependence on the particle spin $s$ for fixed conformal scale $L$, respectively. The lower row shows the dependence on $L$ for fixed particle spins $s$, respectively.}
\label{fig:veff}
\end{figure*}

\subsection{Spinning-particle equations of motion}
\label{section3}

We focus on  the orbiting body. Its motion is governed by the MPD equations~\cite{Mathisson1937,Papapetrou1951,Dixon1970}
\begin{align}
\frac{Dp^\alpha}{d\lambda} &=
-\frac{1}{2}R^\alpha_{\ \beta\delta\sigma}u^\beta S^{\delta\sigma},
\nonumber\\
\frac{DS^{\alpha\beta}}{d\lambda} &=
p^\alpha u^\beta-p^\beta u^\alpha ,
\label{MPD}
\end{align}
where $p^\mu$ is the dynamical four-momentum, $u^\mu=dx^\mu/d\lambda$ is the tangent vector to the representative worldline, and $S^{\mu\nu}$ is the antisymmetric spin tensor. We impose the Tulczyjew condition~\cite{Tulczyjew1959}
\begin{equation}
S^{\alpha\beta}p_\alpha=0,
\label{eq:ssc}
\end{equation}
under which the dynamical mass and spin magnitude are conserved~\cite{Tulczyjew1959}
\begin{equation}
p^\alpha p_\alpha=-m^2,\qquad
S^{\alpha\beta}S_{\alpha\beta}=2S^2 .
\label{eq:mass_spin}
\end{equation}
The specific spin used in the numerical analysis is $s=S/m$.

It is useful to rewrite Eq.~\eqref{metric} in the generic spherical form
\begin{equation}
ds^2=-A(r)dt^2+B(r)dr^2+C(r)d\Omega^2,
\label{eq:ABCmetric}
\end{equation}
with
\begin{equation}
A(r)=S(r)f(r),\qquad B(r)=\frac{S(r)}{f(r)},\qquad C(r)=S(r)r^2 .
\label{eq:ABC}
\end{equation}
For equatorial motion, $\theta=\pi/2$, and spin orthogonal to the orbital plane, the relevant spin-tensor component can be written as~\cite{Witzany2019,Plyatsko2011}
\begin{equation}
S^{tr}=\frac{s\,p_\varphi}{\sqrt{A(r)B(r)C(r)}}.
\label{eq:Str}
\end{equation}
The Killing vectors $\partial_t$ and $\partial_\varphi$ then provide two conserved quantities,
\begin{align}
E &=
-p_t-\alpha(r)s\,p_\varphi,
\nonumber\\
l &=
p_\varphi+\beta(r)s\,p_t,
\label{eq:EJ}
\end{align}
where $E$ is the specific energy, $l$ is the total angular momentum, and
\begin{equation}
\alpha(r)=\frac{A'(r)}{2\sqrt{A(r)B(r)C(r)}},
\qquad
\beta(r)=\frac{C'(r)}{2\sqrt{A(r)B(r)C(r)}} .
\label{eq:alphabeta}
\end{equation}
Solving Eq.~\eqref{eq:EJ} gives
\begin{align}
p_t &=
-\frac{E+\alpha(r)sl}{1-\alpha(r)\beta(r)s^2},
\nonumber\\
p_\varphi &=
\frac{l+\beta(r)sE}{1-\alpha(r)\beta(r)s^2}.
\label{eq:ptpphi}
\end{align}
Thus the conformal scale modifies the dynamics through $A(r)$, $B(r)$, $C(r)$ and their derivatives, not only through the Schwarzschild factor $f(r)$.

The radial equation follows from the mass-shell condition,
\begin{equation}
p_r^2
=B(r)\left[
\frac{p_t^2}{A(r)}
-\frac{p_\varphi^2}{C(r)}
-m^2
\right].
\label{eq:pr2}
\end{equation}
After substituting Eq.~\eqref{eq:ptpphi}, Eq.~\eqref{eq:pr2} becomes a quadratic equation in $E$ and can be factorized as
\begin{equation}
p_r^2=
\left(E-V_{\rm eff}^{+}\right)
\left(E-V_{\rm eff}^{-}\right),
\label{eq:effective}
\end{equation}
where $V_{\rm eff}^{+}$ is the positive-energy branch. This branch is used to determine circular orbits, the ISCO and the radial turning points of the trajectories.

Figure~\ref{fig:veff} shows the effective potential for representative values of $s$ and $L$. In the Schwarzschild limit $L=0$, positive spin raises the potential barrier whereas negative spin lowers it, showing the direct influence of spin--curvature coupling. When $L$ is increased at fixed spin, the peak of the potential becomes higher and shifts in radius. The conformal factor therefore strengthens the near-horizon effective barrier and changes the location of stable and unstable circular orbits. This behavior is the first indication that $L$ and $s$ affect the orbital dynamics in different but coupled ways.

\begin{figure*}
\centering
\includegraphics[width=0.45\textwidth]{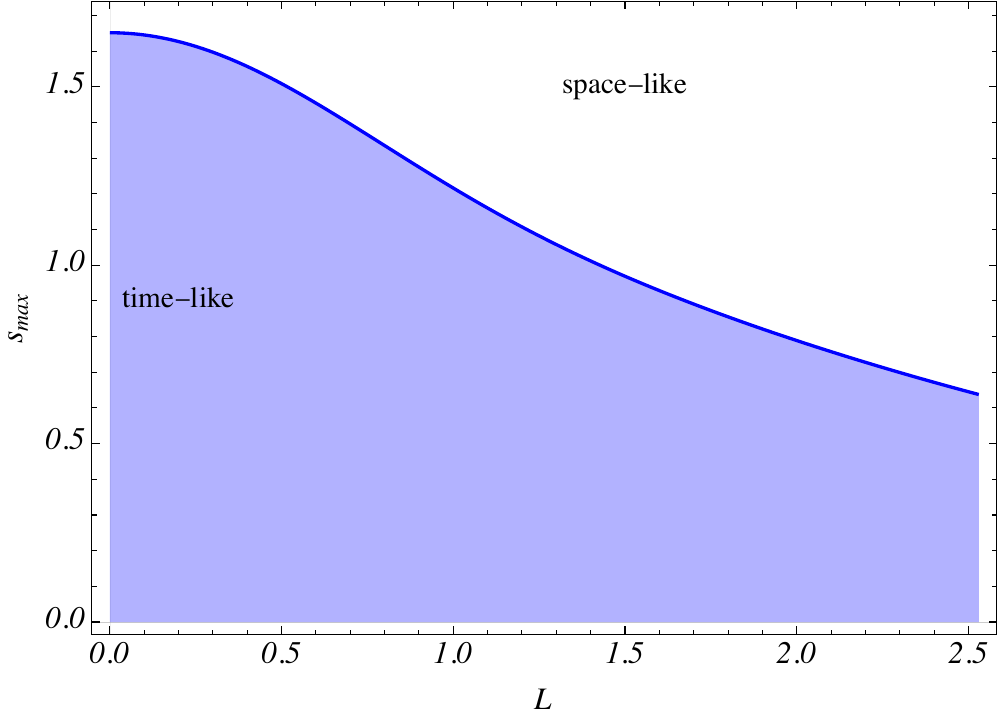}\hfill
\includegraphics[width=0.5\textwidth]{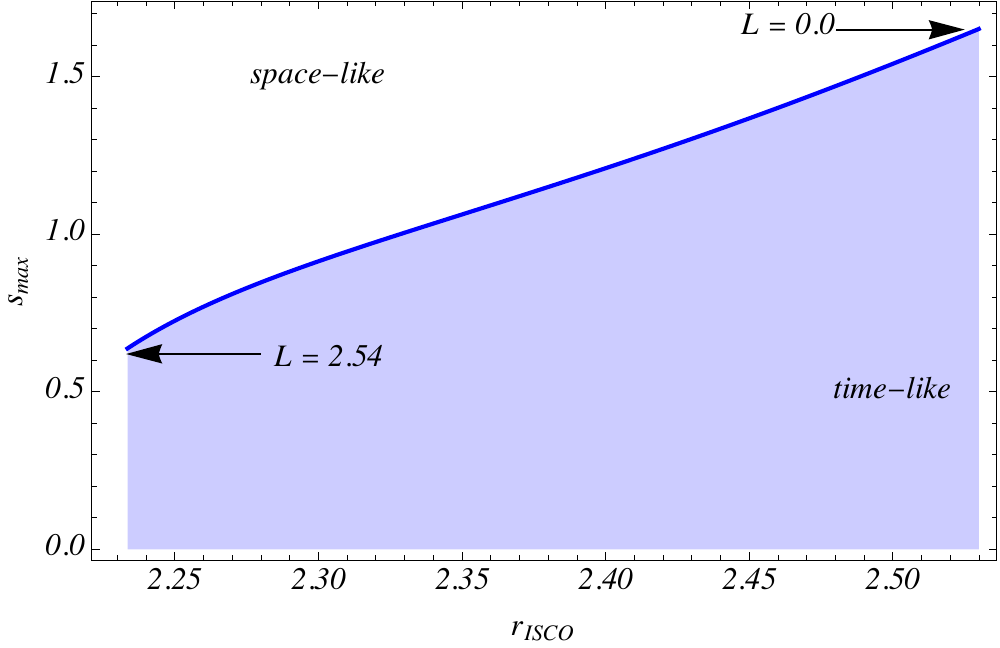}
\caption{Maximum admissible dimensionless spin $s_{\rm max}$ obtained from the timelike-motion condition. The left panel shows $s_{\rm max}$ as a function of the conformal scale $L$. The right panel shows the same bound as a function of the ISCO radius; the curve is parameterized by $L$, with the Schwarzschild limit $L=0$ at the upper-right end and the largest plotted conformal scale near $L=2.54$ at the lower-left end. The shaded region is timelike and physically admissible, while the region above the curve is spacelike and excluded.}
\label{fig:super_main}
\end{figure*}

\begin{figure*}
\centering
\includegraphics[width=0.325\textwidth]{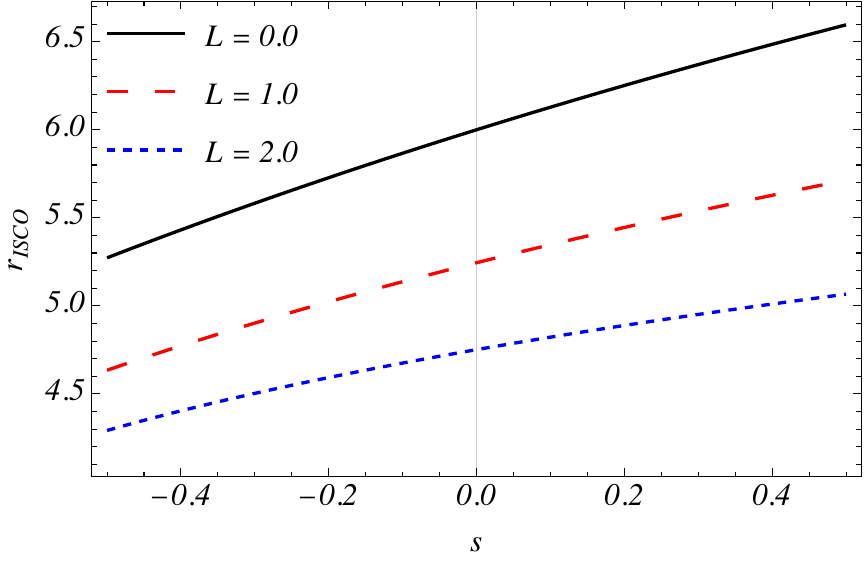}
\includegraphics[width=0.325\textwidth]{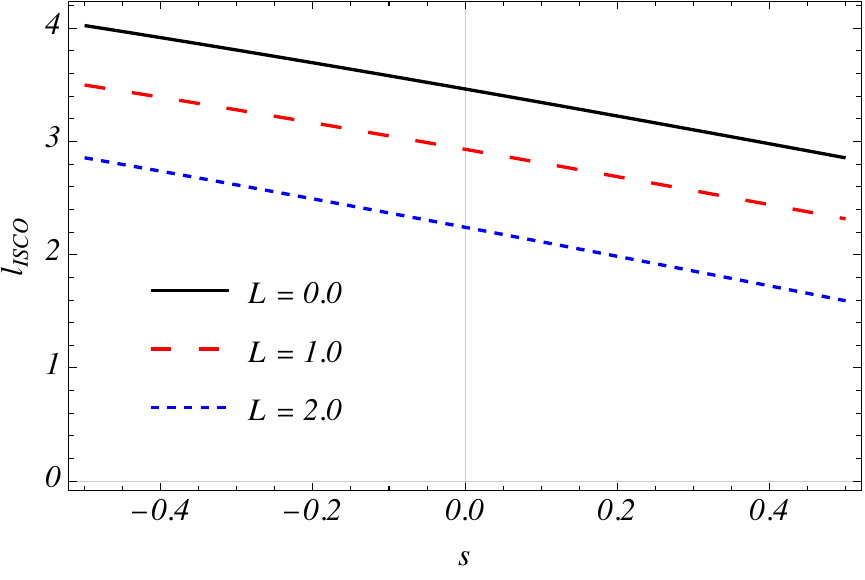}
\includegraphics[width=0.325\textwidth]{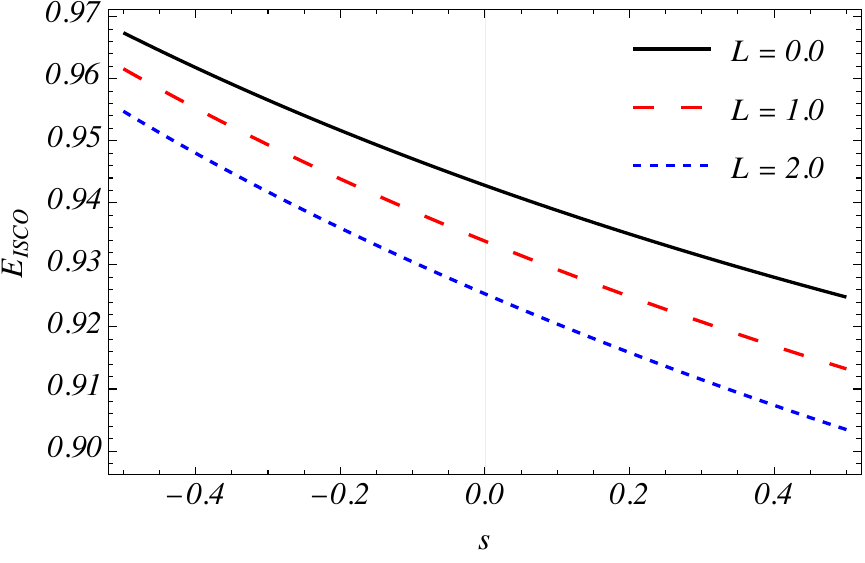}\\[0.5em]
\caption{ISCO properties of spinning particles in the conformal Schwarzschild spacetime. The left, middle and right panels show the ISCO radius $r_{\rm ISCO}$, angular momentum $l_{\rm ISCO}$ and energy $E_{\rm ISCO}$ as functions of the particle spin $s$.}
\label{fig:isco}
\end{figure*}

\subsection{Timelike bound}
\label{section4}

Under the Tulczyjew condition the four-momentum is timelike, but the four-velocity is not generally parallel to it. For sufficiently large spin, the MPD velocity can become spacelike even when $p_\mu p^\mu=-m^2$ remains satisfied \cite{Saijo1998,Hartl2003}. Such trajectories are unphysical. We therefore impose the timelike condition
\begin{equation}
\frac{u_\mu u^\mu}{(u^t)^2}
=g_{tt}
+g_{rr}\left(\frac{dr}{dt}\right)^2
+g_{\phi\phi}\left(\frac{d\phi}{dt}\right)^2
\leq0 .
\label{eq:timelike_condition}
\end{equation}
The equality defines the limiting spin $s_{\rm max}$ for each value of the conformal scale.

Figure~\ref{fig:super_main} presents the resulting physical-spin domain. The left panel shows that $s_{\rm max}$ decreases monotonically as $L$ grows over the plotted interval. Hence the conformal deformation makes the  timelike condition more restrictive: a particle spin that is admissible in the Schwarzschild limit can become superluminal when the conformal scale is sufficiently large. The right panel translates the same bound into the ISCO plane. Since increasing $L$ decreases $r_{\rm ISCO}$, the allowed spin becomes smaller for more compact marginally stable orbits. The shaded region below the curve identifies the domain used for physically meaningful MPD trajectories.

This restriction is also a consistency condition for the approximation. In astrophysical extreme-mass-ratio systems one expects $s\sim\chi m/M$, where $\chi\leq1$ is the dimensionless spin of the small body; hence realistic test bodies usually satisfy $|s|\ll1$ \cite{Hartl2003}. The comparatively larger values used in the figures are useful for displaying the structure of the timelike boundary and for isolating the impact of the conformal scale.

\subsection{Innermost stable circular orbit}
\label{sec:isco}

Circular equatorial motion is determined by the extrema of the positive branch of the effective potential. The circular-orbit conditions are~\cite{AaAbdulxamidov2022,Turakhonov2026}
\begin{equation}
E=V_{\rm eff}^{+},\qquad
\frac{dV_{\rm eff}^{+}}{dr}=0,
\label{eq:circular}
\end{equation}
and stability requires
\begin{equation}
\frac{d^2V_{\rm eff}^{+}}{dr^2}\geq0 .
\label{eq:stability}
\end{equation}
The ISCO is obtained when the equality in Eq.~\eqref{eq:stability} holds. We solve these equations numerically for $r_{\rm ISCO}$, $l_{\rm ISCO}$ and $E_{\rm ISCO}$.

Figure~\ref{fig:isco} summarizes the ISCO behavior. For each fixed value of $L$, increasing the particle spin shifts the ISCO outward. At the same time, $l_{\rm ISCO}$ and $E_{\rm ISCO}$ decrease as $s$ increases. The conformal scale acts in the opposite radial direction: for any fixed spin, larger $L$ decreases $r_{\rm ISCO}$ and also lowers both $l_{\rm ISCO}$ and $E_{\rm ISCO}$. The conformal Schwarzschild black hole therefore supports marginally stable circular motion closer to the horizon than the Schwarzschild case, while the spin--curvature coupling can counteract that inward shift when the spin is positive.

\begin{figure*}
\centering
\includegraphics[width=0.32\textwidth]{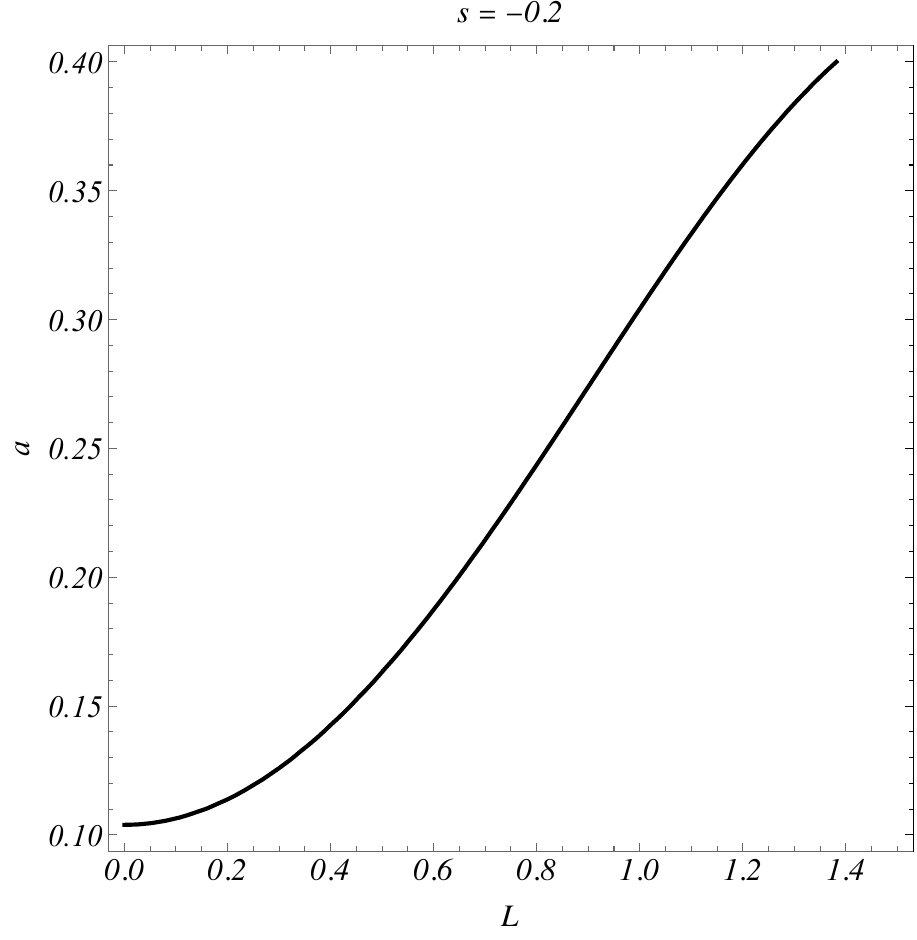}
\includegraphics[width=0.32\textwidth]{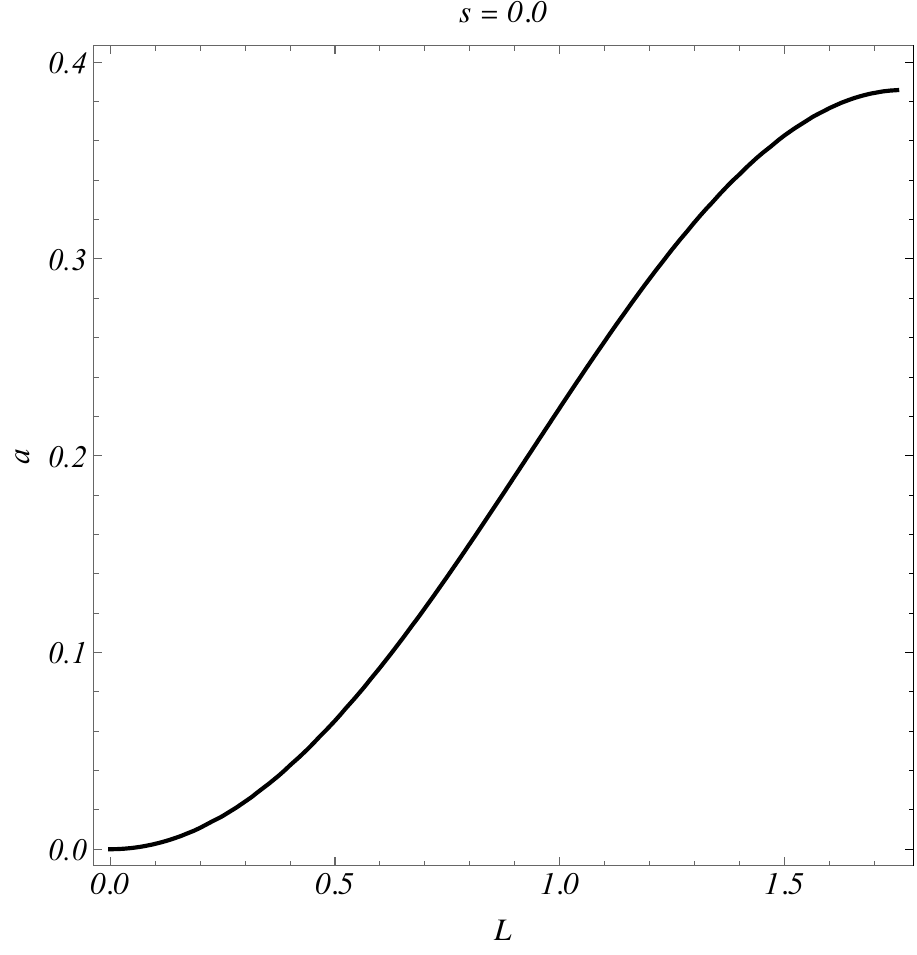}
\includegraphics[width=0.32\textwidth]{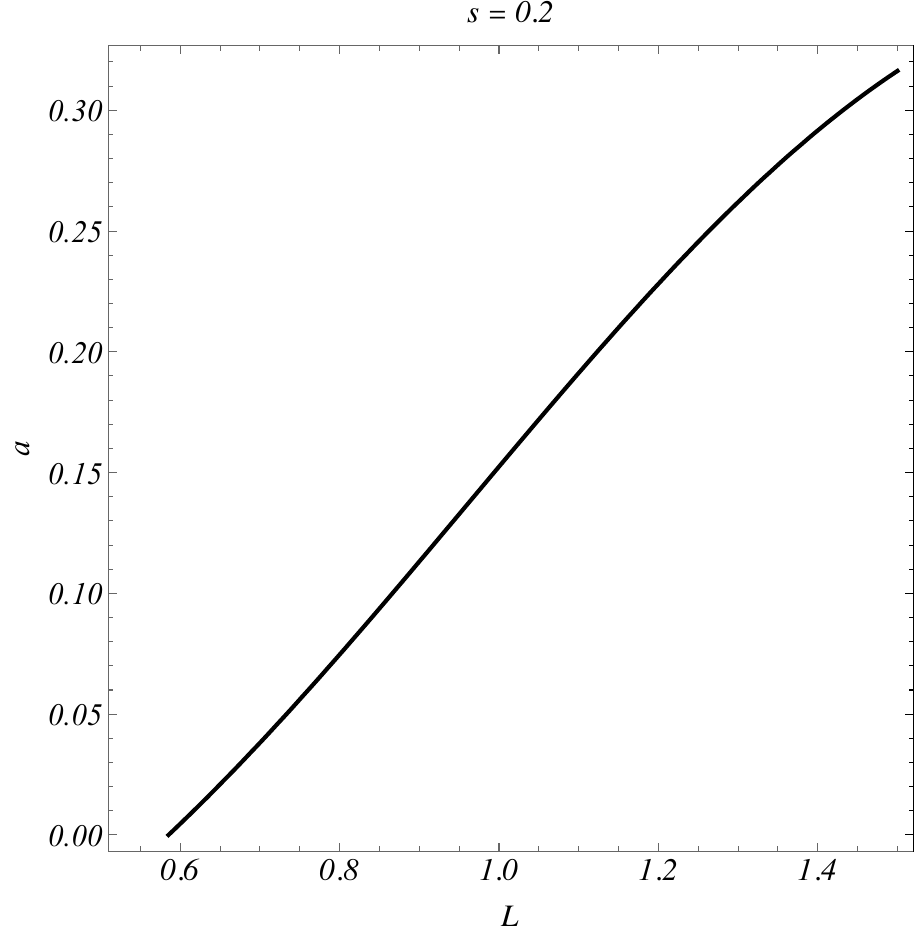}\\[0.5em]
\caption{Degeneracy between the Kerr rotation parameter $a$ and the conformal scale $L$ obtained by matching the ISCO radius of spinning particles. Each curve gives the Kerr spin that produces the same ISCO radius as the conformal Schwarzschild black hole with the indicated value of $L$.}
\label{all:kerr}
\end{figure*}

Figure~\ref{all:kerr} compares the ISCO displacement produced by the conformal scale with the one produced by Kerr rotation. For each fixed particle spin, the required Kerr parameter $a$ increases with $L$. This means that part of the ISCO shift caused by a rotating Kerr black hole can be mimicked by the conformal deformation of a non-rotating Schwarzschild black hole. The slope and range of the degeneracy depend on the particle spin: positive spin reduces the Kerr parameter required to reproduce a given conformal ISCO shift, while negative spin requires a larger Kerr value. Spin effects must therefore be included when ISCO measurements are used to constrain the conformal scale.

\section{Trajectory structure of spinning particle motion}
\label{section6}

We now integrate representative equatorial trajectories to visualize the combined effects of the conformal scale and the particle spin. The numerical integration uses the same MPD effective-potential framework described above. Eliminating the coordinate time, the orbital curve can be represented through~\cite{Umarov2025,Turakhonov2026}
\begin{equation}
\frac{dr}{d\phi}=\frac{\dot r}{\dot\phi},
\label{eq:traj1}
\end{equation}
and projected onto the equatorial plane as
\begin{equation}
x=r\cos\phi,\qquad
y=r\sin\phi .
\label{eq:traj2}
\end{equation}
The conserved quantities and initial position are kept fixed while $L$ and $s$ are varied, so that the influence of the conformal geometry can be separated from the spin--curvature interaction.

\begin{figure*}
\centering
\includegraphics[width=0.31\textwidth]{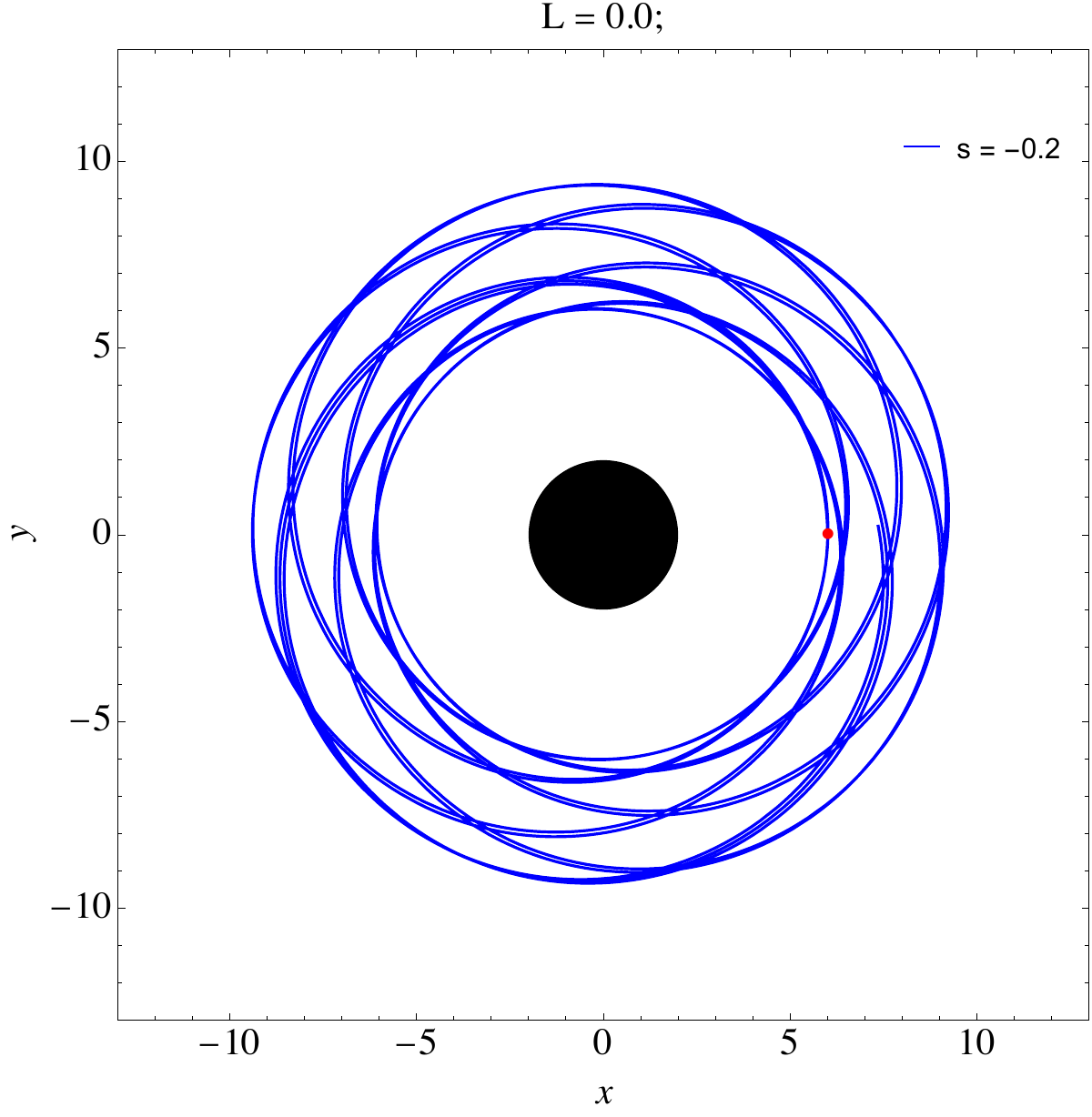}
\includegraphics[width=0.31\textwidth]{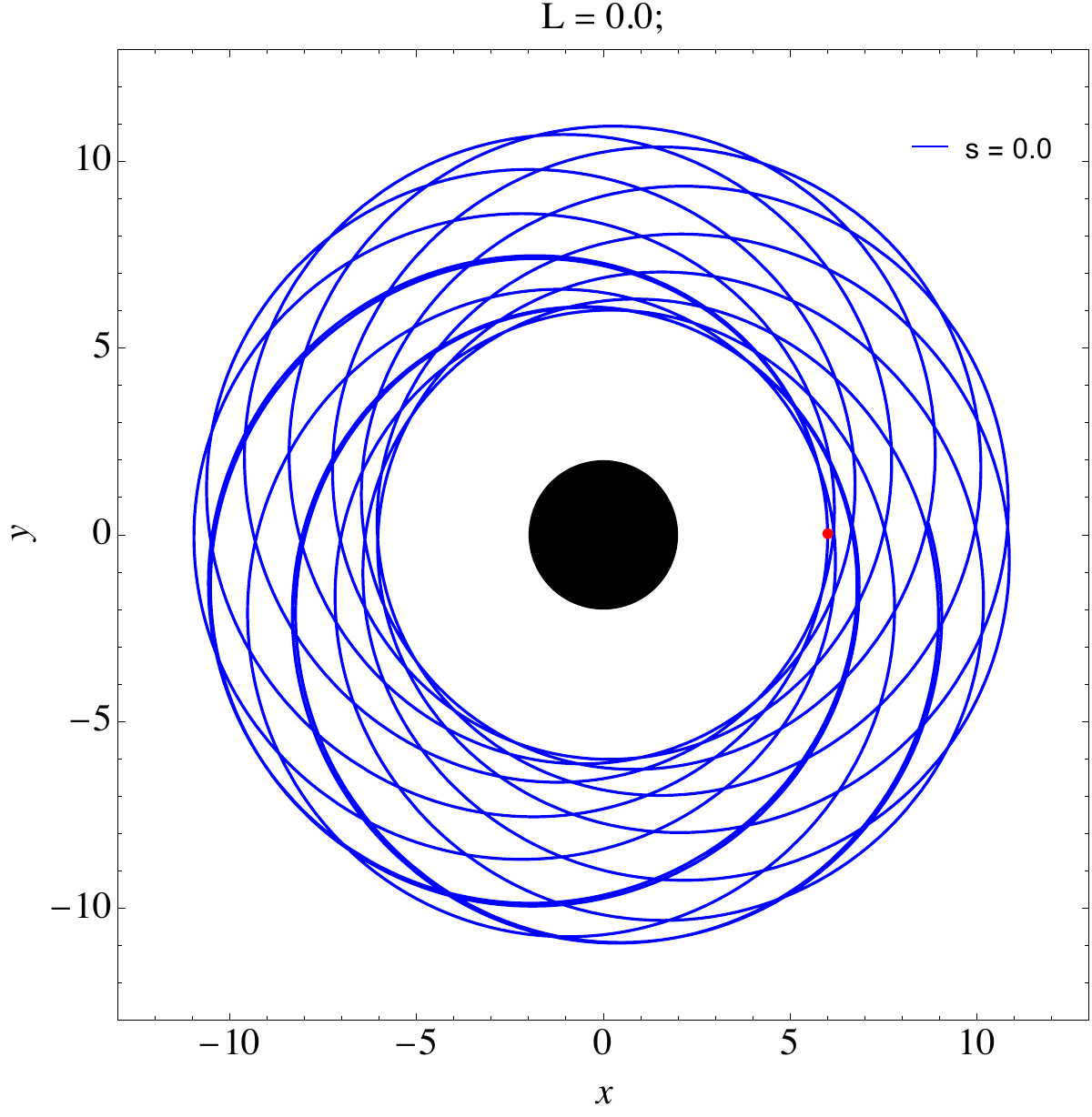}
\includegraphics[width=0.31\textwidth]{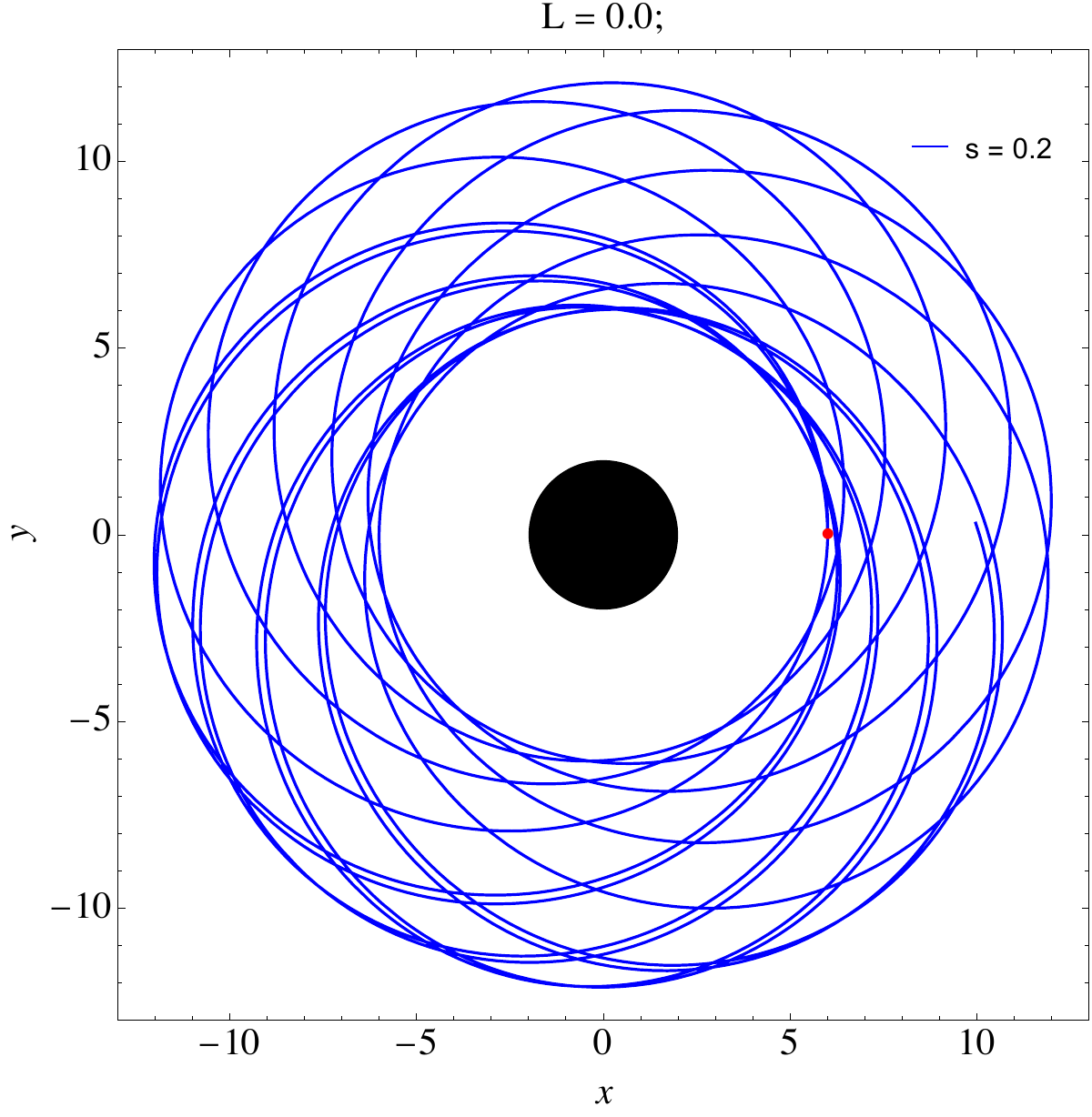}\\[0.3em]
\includegraphics[width=0.31\textwidth]{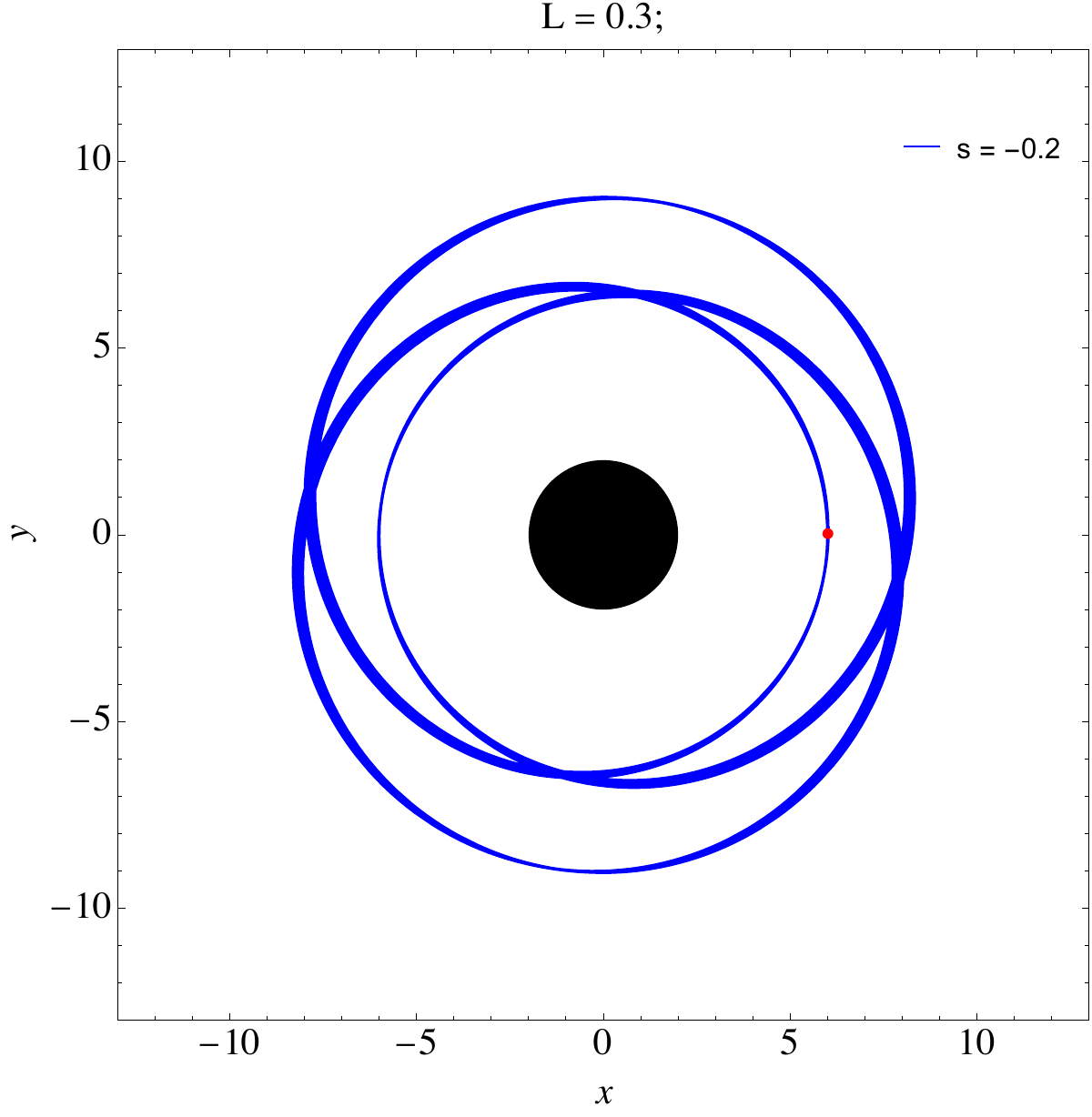}
\includegraphics[width=0.31\textwidth]{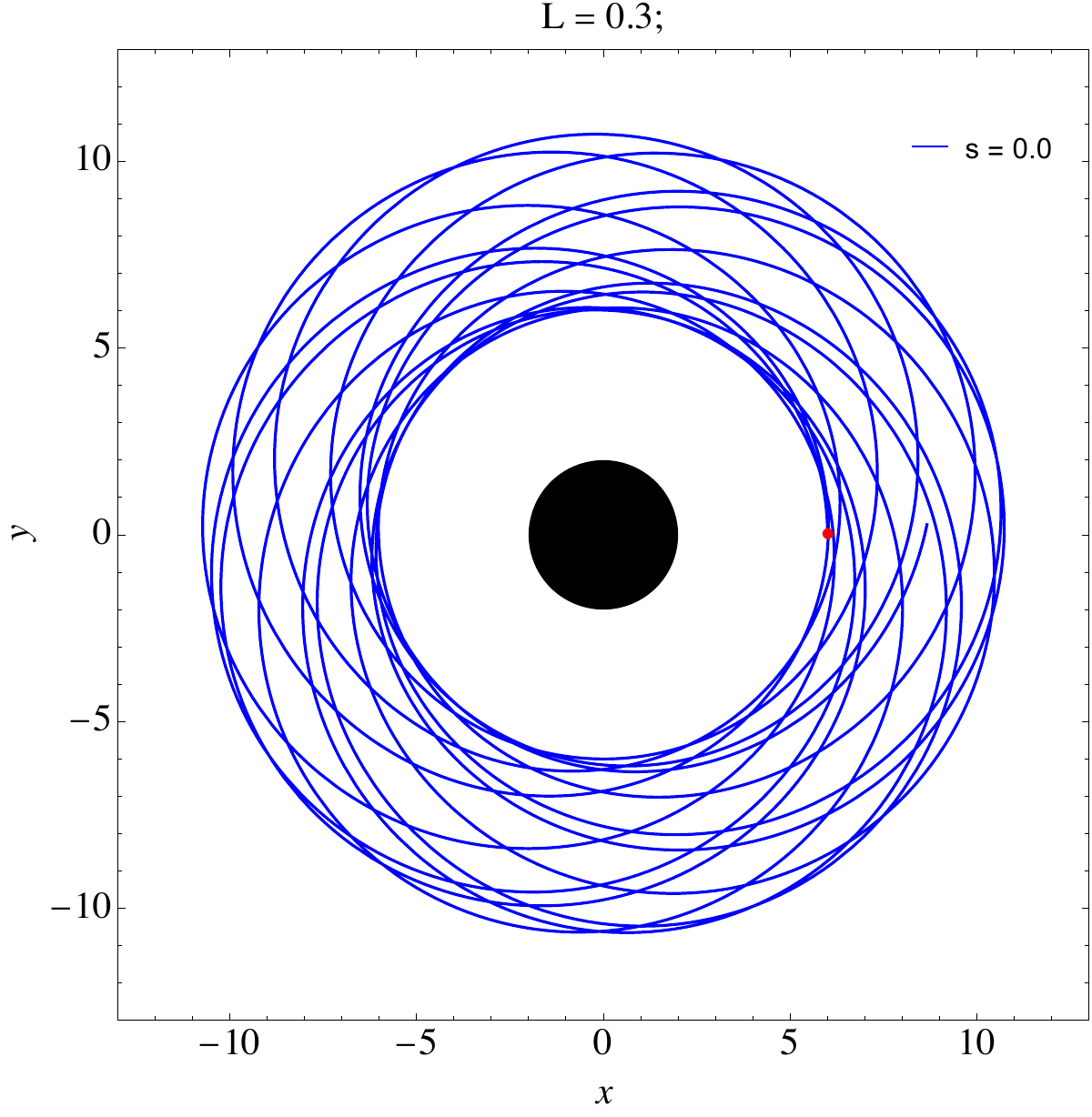}
\includegraphics[width=0.31\textwidth]{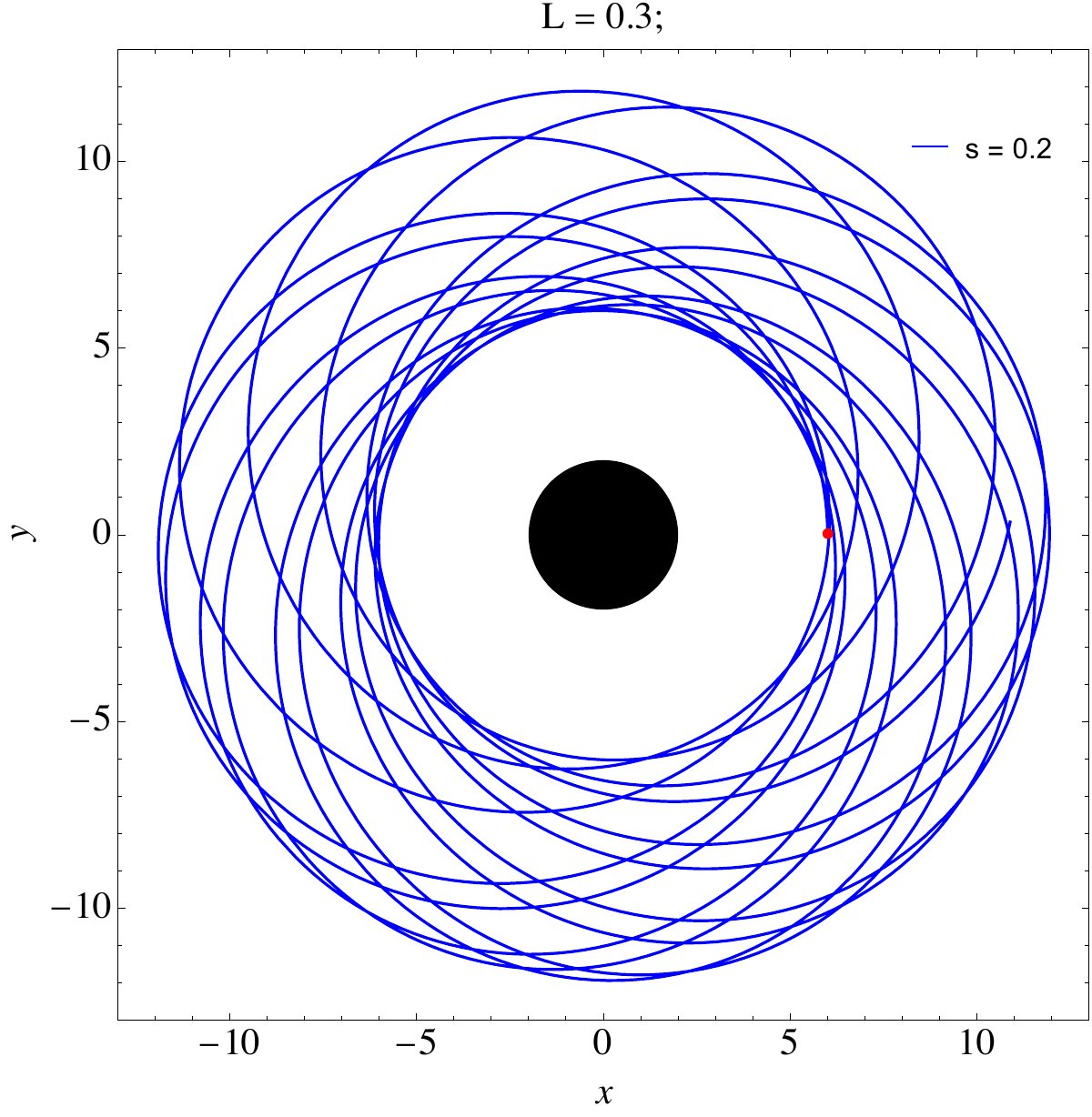}\\[0.3em]
\includegraphics[width=0.31\textwidth]{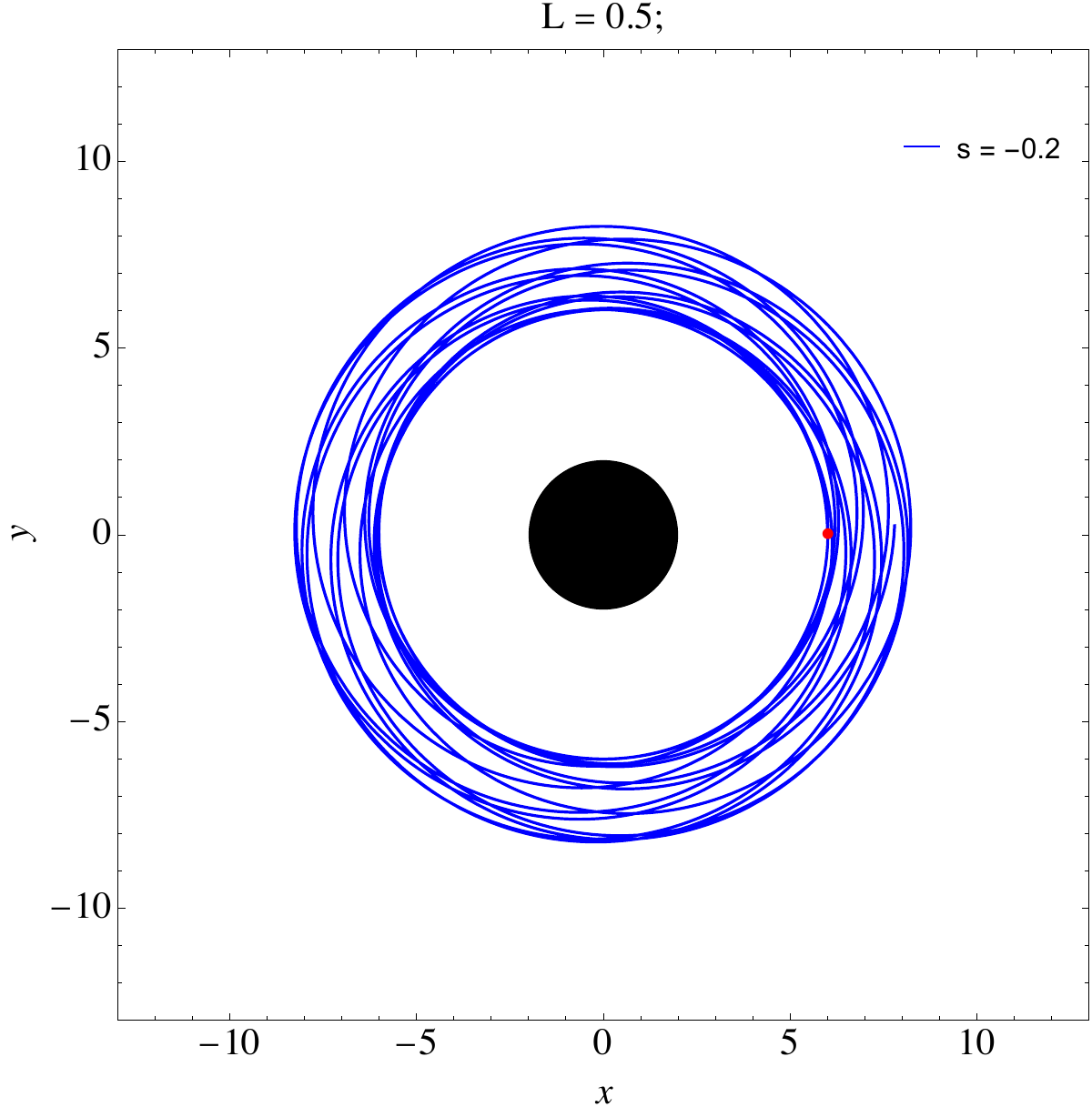}
\includegraphics[width=0.31\textwidth]{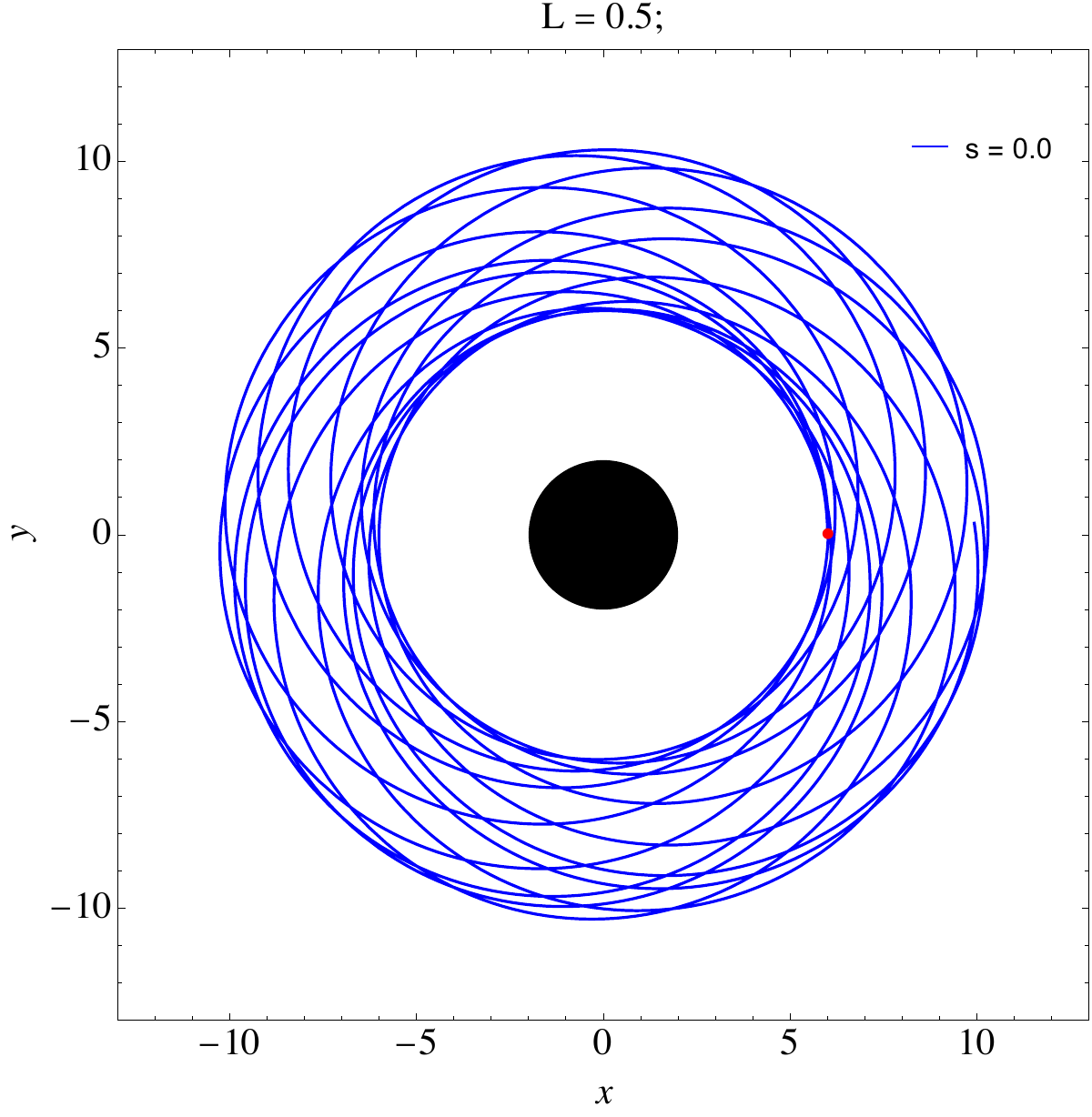}
\includegraphics[width=0.31\textwidth]{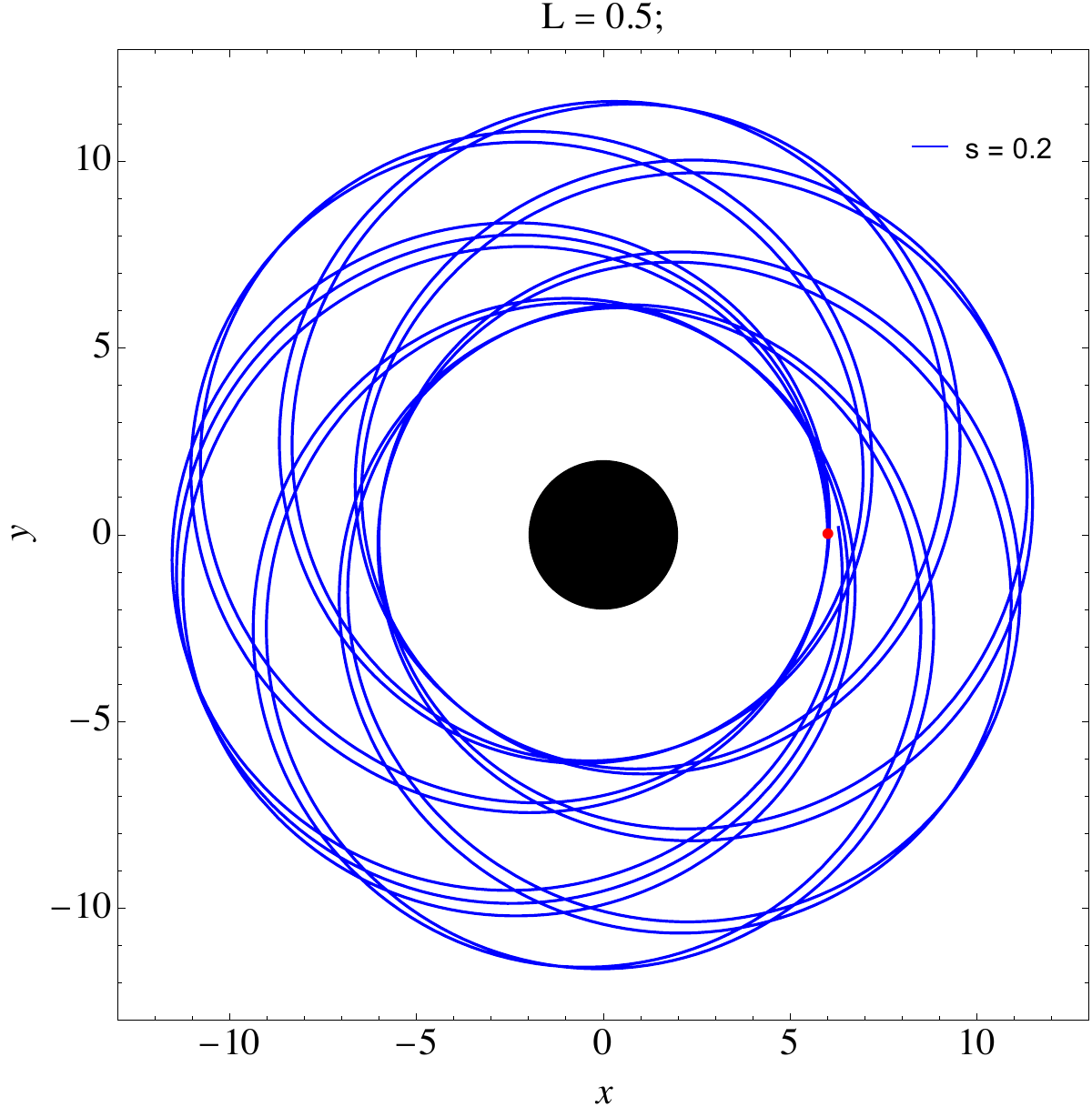}
\caption{Equatorial trajectories of spinning particles around the conformal Schwarzschild black hole. The black disk denotes the horizon and the red point marks the initial position. }
\label{fig:traj1}
\end{figure*}

Figure~\ref{fig:traj1} shows that the conformal scale primarily controls the size of the bound orbit. In the Schwarzschild limit $L=0$, the trajectories occupy a wider radial domain. When $L$ is increased, the orbit contracts and the radial turning points move inward, in agreement with the ISCO trend in Fig.~\ref{fig:isco}. This compactification is a direct consequence of the conformal factor in Eq.~\eqref{metric_function}, which modifies the radial effective potential without moving the event horizon.

The particle spin mainly affects the precession pattern. For a fixed conformal scale, positive spin produces a larger accumulated azimuthal advance than negative spin, while the spinless case lies between them. This reflects the sign-sensitive spin--curvature force in the MPD equations. The trajectory plots therefore complement the effective-potential and ISCO analyses: $L$ changes the radial scale of the orbit, whereas $s$ controls how the orbit winds around the black hole.

\section{Conclusions}
\label{conclusion}

In this work, we investigated the dynamics of spinning test particles around a non-singular Schwarzschild black hole in conformal gravity. The background geometry was described by a conformally rescaled Schwarzschild metric with the conformal factor $S(r)$, where the parameter $L$ controls the deviation from the standard Schwarzschild spacetime. This conformal deformation removes the curvature singularity at $r=0$ while preserving the event horizon at $r=2M$, thereby providing a regular and geodesically complete black-hole geometry.

Using the Mathisson--Papapetrou--Dixon equations together with the Tulczyjew spin supplementary condition, we derived the conserved quantities and constructed the effective potential governing equatorial spinning-particle motion. Our results show that both the particle spin $s$ and the conformal parameter $L$ significantly modify the effective potential. Increasing $L$ enhances the potential barrier and changes the position of its extrema, while the particle spin produces a sign-dependent correction through spin--curvature coupling. These effects directly influence the allowed circular orbits and the radial turning points of the motion.

We also analyzed the physical timelike constraint on the MPD trajectories. The results demonstrate that the maximum admissible spin decreases as the conformal parameter grows, indicating that the conformal deformation strengthens the restriction imposed by the timelike-motion condition. Thus, the parameter $L$ not only modifies the orbital structure but also narrows the physically allowed spin domain for pole--dipole particles.

The ISCO analysis reveals one of the main results of this study. For fixed particle spin, increasing the conformal parameter $L$ decreases the ISCO radius, angular momentum and energy, allowing marginally stable orbits closer to the horizon than in the Schwarzschild case. In contrast, increasing the particle spin shifts the ISCO outward while reducing the corresponding energy and angular momentum. We further showed that the ISCO shift caused by the conformal parameter can mimic the effect of the Kerr rotation parameter. This establishes a degeneracy between the conformal deformation and black-hole rotation, which must be taken into account when interpreting strong-field orbital observables.

Finally, the numerical trajectory analysis confirms the behavior inferred from the effective-potential and ISCO studies. The conformal parameter mainly controls the compactness of the orbit, whereas the particle spin governs the accumulated relativistic precession. Therefore, spinning-particle trajectories provide a sensitive probe of the geometry and can help distinguish conformally regular black holes from standard Schwarzschild and Kerr black holes.

In general, the results show that spin--curvature coupling is an effective diagnostic tool for testing non-singular black holes in conformal gravity. The conformal parameter leaves clear imprints on the effective potential, timelike bound, ISCO properties, Kerr degeneracy and orbital morphology. These findings suggest that future studies of epicyclic frequencies, accretion-disk emission, quasi-periodic oscillations and gravitational-wave signals from spinning compact objects may provide observational constraints on the conformal scale $L$ and offer a possible way to test conformal regularization of black-hole singularities.

\bibliographystyle{apsrev4-1}
\bibliography{1_paper}

@article{Zhang:2018EPJC,
    author = "Zhang, Qiqi and Modesto, Leonardo and Bambi, Cosimo",
    title = "{A general study of regular and singular black hole solutions in Einstein{\textquoteright}s conformal gravity}",
    eprint = "1805.00640",
    archivePrefix = "arXiv",
    primaryClass = "gr-qc",
    doi = "10.1140/epjc/s10052-018-5987-6",
    journal = "Eur. Phys. J. C",
    volume = "78",
    number = "6",
    pages = "506",
    year = "2018"
}

@article{Turimov:2018PRD,
    author = "Turimov, Bobur and Ahmedov, Bobomurat and Abdujabbarov, Ahmadjon and Bambi, Cosimo",
    title = "{Electromagnetic fields of slowly rotating magnetized compact stars in conformal gravity}",
    eprint = "1805.00005",
    archivePrefix = "arXiv",
    primaryClass = "gr-qc",
    doi = "10.1103/PhysRevD.97.124005",
    journal = "Phys. Rev. D",
    volume = "97",
    pages = "124005",
    year = "2018"
}

@article{Bambi:2016JCAPa,
    author = "Bambi, Cosimo and Modesto, Leonardo and Porey, Shiladitya and Rachwa{\l}, Les{\l}aw",
    title = "{Black hole evaporation in conformal gravity}",
    eprint = "1611.05582",
    archivePrefix = "arXiv",
    primaryClass = "gr-qc",
    doi = "10.1088/1475-7516/2017/09/033",
    journal = "JCAP",
    volume = "09",
    pages = "033",
    year = "2017"
}

@article{Bambi:2016JCAP,
    author = "Bambi, Cosimo and Modesto, Leonardo and Rachwa{\l}, Leslaw",
    title = "{Spacetime completeness of non-singular black holes in conformal gravity}",
    eprint = "1611.00865",
    archivePrefix = "arXiv",
    primaryClass = "gr-qc",
    doi = "10.1088/1475-7516/2017/05/003",
    journal = "JCAP",
    volume = "05",
    pages = "003",
    year = "2017"
}

@article{Bambi:2017PRD,
    author = "Bambi, Cosimo and Cao, Zheng and Modesto, Leonardo",
    title = "{Testing conformal gravity with astrophysical black holes}",
    eprint = "1701.00226",
    archivePrefix = "arXiv",
    primaryClass = "gr-qc",
    doi = "10.1103/PhysRevD.95.064006",
    journal = "Phys. Rev. D",
    volume = "95",
    number = "6",
    pages = "064006",
    year = "2017"
}

@article{Plyatsko2011,
  author        = {Plyatsko, Roman and others},
  title         = {Mathisson-Papapetrou-Dixon equations in the Schwarzschild and Kerr backgrounds},
  journal       = {Classical and Quantum Gravity},
  volume        = {28},
  pages         = {195017},
  year          = {2011},
  doi           = {10.1088/0264-9381/28/19/195017},
  eprint        = {1106.xxxxx},
  archivePrefix = {arXiv},
  primaryClass  = {gr-qc}
}

@article{Witzany2019,
  author        = {Witzany, Vojt{\v{e}}ch and Steinhoff, Jan and Lukes-Gerakopoulos, Georgios},
  title         = {Hamiltonians and canonical coordinates for spinning particles in curved space-time},
  journal       = {Classical and Quantum Gravity},
  volume        = {36},
  pages         = {075003},
  year          = {2019},
  doi           = {10.1088/1361-6382/ab03a8},
  eprint        = {1808.xxxxx},
  archivePrefix = {arXiv},
  primaryClass  = {gr-qc}
}

@article{EHT2024Horndeski,
  author        = {Kumar, Rahul and others},
  title         = {Shadows and quasinormal modes of rotating black holes in Horndeski theory: Parameter constraints using EHT observations of M87* and Sgr A*},
  journal       = {Physics of the Dark Universe},
  volume        = {46},
  pages         = {101560},
  year          = {2024},
  doi           = {10.1016/j.dark.2024.101560},
  eprint        = {2405.xxxxx},
  archivePrefix = {arXiv},
  primaryClass  = {gr-qc}
}

@article{Drummond2023,
  author        = {Drummond, Lisa V. and Lynch, Philip and others},
  title         = {Extreme mass-ratio inspiral and waveforms for a spinning body into a Kerr black hole via osculating geodesics and near-identity transformations},
  journal       = {Physical Review D},
  volume        = {108},
  pages         = {084015},
  year          = {2023},
  doi           = {10.1103/PhysRevD.108.084015},
  eprint        = {2304.xxxxx},
  archivePrefix = {arXiv},
  primaryClass  = {gr-qc}
}

@article{BenavidesGallego2022,
  author        = {Benavides-Gallego, C. A. and others},
  title         = {Spinning test body orbiting around a Kerr black hole: Comparing spin supplementary conditions for circular equatorial orbits},
  journal       = {Physical Review D},
  volume        = {106},
  pages         = {044039},
  year          = {2022},
  doi           = {10.1103/PhysRevD.106.044039},
  eprint        = {2206.xxxxx},
  archivePrefix = {arXiv},
  primaryClass  = {gr-qc}
}

@article{Jefremov2022,
  author        = {Jefremov, Pavel I. and others},
  title         = {Spin-induced changes in the parameters of ISCO in Kerr spacetime},
  journal       = {International Journal of Modern Physics D},
  volume        = {31},
  pages         = {2250012},
  year          = {2022},
  doi           = {10.1142/S0218271822500123}
}

@article{CarballoRubio2023,
  author        = {Carballo-Rubio, Ra\'ul and Di Filippo, Francesco and Liberati, Stefano and Visser, Matt and others},
  title         = {Regular black holes: A short topic review},
  journal       = {International Journal of Theoretical Physics},
  volume        = {62},
  pages         = {185},
  year          = {2023},
  doi           = {10.1007/s10773-023-05454-1},
  eprint        = {2303.xxxxx},
  archivePrefix = {arXiv},
  primaryClass  = {gr-qc}
}

@book{Bambi2023Book,
  editor        = {Bambi, Cosimo},
  title         = {Regular Black Holes: Towards a New Paradigm of Gravitational Collapse},
  publisher     = {Springer},
  year          = {2023},
  series        = {Springer Series in Astrophysics and Cosmology},
  doi           = {10.1007/978-981-99-1596-5}
}

@article{Wei2024,
  author        = {Wei, Shao-Wen and others},
  title         = {Images of nonsingular nonrotating black holes in conformal gravity},
  journal       = {Physics Letters B},
  volume        = {850},
  pages         = {138534},
  year          = {2024},
  doi           = {10.1016/j.physletb.2024.138534},
  eprint        = {2312.xxxxx},
  archivePrefix = {arXiv},
  primaryClass  = {gr-qc}
}

@article{Bambi2019QNM,
  author        = {Bambi, Cosimo and Modesto, Leonardo and Porey, Shiladitya and Rachwa{\l}, Les{\l}aw},
  title         = {Gravitational perturbations of nonsingular black holes in conformal gravity},
  journal       = {Physical Review D},
  volume        = {99},
  pages         = {104044},
  year          = {2019},
  doi           = {10.1103/PhysRevD.99.104044},
  eprint        = {1903.xxxxx},
  archivePrefix = {arXiv},
  primaryClass  = {gr-qc}
}

@article{Zhang2023Conformal,
  author        = {Zhang, Yu and others},
  title         = {Black hole shadows in fourth-order conformal Weyl gravity},
  journal       = {European Physical Journal C},
  volume        = {83},
  pages         = {11274},
  year          = {2023},
  doi           = {10.1140/epjc/s10052-023-11274-8}
}

@article{Cai2024,
  author        = {Cai, Xiao-Chang and Miao, Yan-Gang},
  title         = {Quasinormal modes and shadows of a new family of Ay\'on-Beato--Garc\'ia black holes},
  journal       = {Physical Review D},
  volume       = {109},
  pages         = {084013},
  year          = {2024},
  doi           = {10.1103/PhysRevD.109.084013},
  eprint        = {2312.xxxxx},
  archivePrefix = {arXiv},
  primaryClass  = {gr-qc}
}

@article{ConformalThermo2022,
  author        = {Singh, D. V. and others},
  title         = {Study of thermodynamical geometries of conformal gravity black hole},
  journal       = {European Physical Journal C},
  volume        = {82},
  pages         = {1029},
  year          = {2022},
  doi           = {10.1140/epjc/s10052-022-10956-2}
}

@inproceedings{Bardeen1968,
  author    = {Bardeen, J. M.},
  title     = {Non-singular general-relativistic gravitational collapse},
  booktitle = {Proceedings of the International Conference GR5},
  year      = {1968},
  pages     = {174},
  address   = {Tbilisi, USSR},
  publisher = {Tbilisi University Press}
}

@article{Hayward2006,
  author        = {Hayward, Sean A.},
  title         = {Formation and evaporation of nonsingular black holes},
  journal       = {Physical Review Letters},
  volume        = {96},
  pages         = {031103},
  year          = {2006},
  eprint        = {gr-qc/0506126},
  archivePrefix = {arXiv},
  primaryClass  = {gr-qc},
  doi           = {10.1103/PhysRevLett.96.031103}
}

@article{Dymnikova1992,
  author  = {Dymnikova, I.},
  title   = {Vacuum nonsingular black hole},
  journal = {General Relativity and Gravitation},
  volume  = {24},
  pages   = {235},
  year    = {1992},
  doi     = {10.1007/BF00760226}
}

@article{AyonBeato1998,
  author        = {Ay\'on-Beato, Eloy and Garc\'ia, Alberto},
  title         = {Regular black hole in general relativity coupled to nonlinear electrodynamics},
  journal       = {Physical Review Letters},
  volume        = {80},
  pages         = {5056},
  year          = {1998},
  eprint        = {gr-qc/9911046},
  archivePrefix = {arXiv},
  primaryClass  = {gr-qc},
  doi           = {10.1103/PhysRevLett.80.5056}
}

@article{Simpson2019,
  author        = {Simpson, Alex and Visser, Matt},
  title         = {Black-bounce to traversable wormhole},
  journal       = {Journal of Cosmology and Astroparticle Physics},
  volume        = {2019},
  number        = {02},
  pages         = {042},
  year          = {2019},
  eprint        = {1812.07114},
  archivePrefix = {arXiv},
  primaryClass  = {gr-qc},
  doi           = {10.1088/1475-7516/2019/02/042}
}

@article{CarballoRubio2018,
  author        = {Carballo-Rubio, Ra\'ul and Di Filippo, Francesco and Liberati, Stefano and Visser, Matt},
  title         = {Regular black holes: An effective framework for non-singular black holes},
  journal       = {Journal of High Energy Physics},
  volume        = {2018},
  number        = {10},
  pages         = {175},
  year          = {2018},
  eprint        = {1806.06367},
  archivePrefix = {arXiv},
  primaryClass  = {gr-qc},
  doi           = {10.1007/JHEP10(2018)175}
}

@article{BenavidesGallego2019,
  author        = {Benavides-Gallego, C. A. and Abdujabbarov, A. and Malafarina, D. and Ahmedov, B. and Bambi, C.},
  title         = {Spinning test particle motion around a rotating regular black hole},
  journal       = {Physical Review D},
  volume        = {99},
  pages         = {044012},
  year          = {2019},
  eprint        = {1812.04846},
  archivePrefix = {arXiv},
  primaryClass  = {gr-qc},
  doi           = {10.1103/PhysRevD.99.044012}
}

@article{Cardoso2019,
  author        = {Cardoso, Vitor and Pani, Paolo},
  title         = {Testing the nature of dark compact objects: a status report},
  journal       = {Living Reviews in Relativity},
  volume        = {22},
  pages         = {4},
  year          = {2019},
  eprint        = {1904.05363},
  archivePrefix = {arXiv},
  primaryClass  = {gr-qc},
  doi           = {10.1007/s41114-019-0020-4}
}

@article{Barack2019,
  author        = {Barack, Leor and others},
  title         = {Black holes, gravitational waves and fundamental physics: a roadmap},
  journal       = {Classical and Quantum Gravity},
  volume        = {36},
  pages         = {143001},
  year          = {2019},
  eprint        = {1806.05195},
  archivePrefix = {arXiv},
  primaryClass  = {gr-qc},
  doi           = {10.1088/1361-6382/ab0587}
}

@article{Will2014,
  author  = {Will, Clifford M.},
  title   = {The Confrontation between General Relativity and Experiment},
  journal = {Living Reviews in Relativity},
  volume  = {17},
  pages   = {4},
  year    = {2014},
  doi     = {10.12942/lrr-2014-4}
}

@article{Berti2015,
  author  = {Berti, Emanuele and Barausse, Enrico and Cardoso, Vitor and others},
  title   = {Testing General Relativity with Present and Future Astrophysical Observations},
  journal = {Classical and Quantum Gravity},
  volume  = {32},
  pages   = {243001},
  year    = {2015},
  doi     = {10.1088/0264-9381/32/24/243001}
}

@article{Abbott2016,
  author  = {{LIGO Scientific Collaboration and Virgo Collaboration}},
  title   = {Observation of Gravitational Waves from a Binary Black Hole Merger},
  journal = {Physical Review Letters},
  volume  = {116},
  pages   = {061102},
  year    = {2016},
  doi     = {10.1103/PhysRevLett.116.061102}
}

@article{EHT2019,
  author  = {{Event Horizon Telescope Collaboration}},
  title   = {First {M87} Event Horizon Telescope Results. {I}. The Shadow of the Supermassive Black Hole},
  journal = {Astrophysical Journal Letters},
  volume  = {875},
  pages   = {L1},
  year    = {2019},
  doi     = {10.3847/2041-8213/ab0ec7}
}

@article{EHT2022,
  author  = {{Event Horizon Telescope Collaboration}},
  title   = {First Sagittarius {A*} Event Horizon Telescope Results. {I}. The Shadow of the Supermassive Black Hole in the Center of the Milky Way},
  journal = {Astrophysical Journal Letters},
  volume  = {930},
  pages   = {L12},
  year    = {2022},
  doi     = {10.3847/2041-8213/ac6674}
}

@article{Penrose1965,
  author  = {Penrose, Roger},
  title   = {Gravitational Collapse and Space-Time Singularities},
  journal = {Physical Review Letters},
  volume  = {14},
  pages   = {57--59},
  year    = {1965},
  doi     = {10.1103/PhysRevLett.14.57}
}

@article{HawkingPenrose1970,
  author  = {Hawking, Stephen W. and Penrose, Roger},
  title   = {The Singularities of Gravitational Collapse and Cosmology},
  journal = {Proceedings of the Royal Society A},
  volume  = {314},
  pages   = {529--548},
  year    = {1970},
  doi     = {10.1098/rspa.1970.0021}
}

@article{MannheimKazanas1989,
  author  = {Mannheim, Philip D. and Kazanas, Demosthenes},
  title   = {Exact Vacuum Solution to Conformal Weyl Gravity and Galactic Rotation Curves},
  journal = {Astrophysical Journal},
  volume  = {342},
  pages   = {635--638},
  year    = {1989},
  doi     = {10.1086/167623}
}

@article{Dabrowski2009,
  author  = {Dabrowski, M. P. and Garecki, J. and Blaschke, D. B.},
  title   = {Conformal Transformations and Conformal Invariance in Gravitation},
  journal = {Annalen der Physik},
  volume  = {521},
  pages   = {13--32},
  year    = {2009},
  doi     = {10.1002/andp.200810331}
}

@article{Clifton2012,
  author  = {Clifton, Timothy and Ferreira, Pedro G. and Padilla, Antonio and Skordis, Constantinos},
  title   = {Modified Gravity and Cosmology},
  journal = {Physics Reports},
  volume  = {513},
  pages   = {1--189},
  year    = {2012},
  doi     = {10.1016/j.physrep.2012.01.001}
}

@article{BambiModestoRachwal2017,
  author        = {Bambi, Cosimo and Modesto, Leonardo and Rachwa{\l}, Les{\l}aw},
  title         = {Spacetime Completeness of Non-Singular Black Holes in Conformal Gravity},
  journal       = {Journal of Cosmology and Astroparticle Physics},
  volume        = {2017},
  number        = {05},
  pages         = {003},
  year          = {2017},
  doi           = {10.1088/1475-7516/2017/05/003},
  eprint        = {1611.00865},
  archivePrefix = {arXiv},
  primaryClass  = {gr-qc}
}

@article{AaToshmatov2017,
  author  = {Toshmatov, Bobir and Bambi, Cosimo and Ahmedov, Bobomurat and Abdujabbarov, Ahmadjon and Stuchlik, Zdenek},
  title   = {Energy Conditions of Non-Singular Black Hole Spacetimes in Conformal Gravity},
  journal = {European Physical Journal C},
  volume  = {77},
  pages   = {542},
  year    = {2017},
  doi     = {10.1140/epjc/s10052-017-5112-2}
}

@article{Mathisson1937,
  author  = {Mathisson, Myron},
  title   = {Neue Mechanik materieller Systeme},
  journal = {Acta Physica Polonica},
  volume  = {6},
  pages   = {163--200},
  year    = {1937}
}

@article{Papapetrou1951,
  author  = {Papapetrou, Achille},
  title   = {Spinning Test-Particles in General Relativity. {I}},
  journal = {Proceedings of the Royal Society A},
  volume  = {209},
  pages   = {248--258},
  year    = {1951},
  doi     = {10.1098/rspa.1951.0200}
}

@article{Dixon1970,
  author  = {Dixon, W. G.},
  title   = {Dynamics of Extended Bodies in General Relativity. {I}. Momentum and Angular Momentum},
  journal = {Proceedings of the Royal Society A},
  volume  = {314},
  pages   = {499--527},
  year    = {1970},
  doi     = {10.1098/rspa.1970.0020}
}

@article{Tulczyjew1959,
  author  = {Tulczyjew, W.},
  title   = {Motion of Multipole Particles in General Relativity Theory},
  journal = {Acta Physica Polonica},
  volume  = {18},
  pages   = {393--409},
  year    = {1959}
}

@article{Saijo1998,
  author  = {Saijo, Motoyuki and Maeda, Kei-ichi and Shibata, Masaru and Mino, Yasushi},
  title   = {Gravitational Waves from a Spinning Particle Plunging into a Kerr Black Hole},
  journal = {Physical Review D},
  volume  = {58},
  pages   = {064005},
  year    = {1998},
  doi     = {10.1103/PhysRevD.58.064005}
}

@article{Hartl2003,
  author  = {Hartl, Michael D.},
  title   = {Dynamics of Spinning Test Particles in Kerr Spacetime},
  journal = {Physical Review D},
  volume  = {67},
  pages   = {024005},
  year    = {2003},
  doi     = {10.1103/PhysRevD.67.024005}
}

@article{Toshmatov2019,
  author  = {Toshmatov, Bobir and Malafarina, Daniele},
  title   = {Spinning Test Particles in the {$\gamma$} Spacetime},
  journal = {Physical Review D},
  volume  = {100},
  pages   = {104052},
  year    = {2019},
  doi     = {10.1103/PhysRevD.100.104052}
}

@article{WitzanyPiovano2024,
  author  = {Witzany, Vojt{\v e}ch and Piovano, Gabriel A.},
  title   = {Analytic Solutions for the Motion of Spinning Particles near Spherically Symmetric Black Holes and Exotic Compact Objects},
  journal = {Physical Review Letters},
  volume  = {132},
  pages   = {171401},
  year    = {2024},
  doi     = {10.1103/PhysRevLett.132.171401}
}

@article{Jefremov2015,
  author        = {Jefremov, Paul I. and Tsupko, Oleg Yu. and Bisnovatyi-Kogan, Gennady S.},
  title         = {Innermost Stable Circular Orbits of Spinning Test Particles in Schwarzschild and Kerr Space-Times},
  journal       = {Physical Review D},
  volume        = {91},
  pages         = {124030},
  year          = {2015},
  doi           = {10.1103/PhysRevD.91.124030},
  eprint        = {1503.07060},
  archivePrefix = {arXiv},
  primaryClass  = {gr-qc}
}

@article{Harms2016,
  author        = {Harms, Enno and Lukes-Gerakopoulos, Georgios and Bernuzzi, Sebastiano and Nagar, Alessandro},
  title         = {Spinning Test Body Orbiting around a Schwarzschild Black Hole: Circular Dynamics and Gravitational-Wave Fluxes},
  journal       = {Physical Review D},
  volume        = {94},
  pages         = {104010},
  year          = {2016},
  doi           = {10.1103/PhysRevD.94.104010},
  eprint        = {1609.00356},
  archivePrefix = {arXiv},
  primaryClass  = {gr-qc}
}

@article{AaAbdulxamidov2022,
  author  = {Abdulxamidov, Farrux and Benavides-Gallego, Carlos A. and Han, Wen-Biao and Rayimbaev, Javlon and Abdujabbarov, Ahmadjon},
  title   = {Spinning Test Particle Motion around a Rotating Wormhole},
  journal = {Physical Review D},
  volume  = {106},
  pages   = {024012},
  year    = {2022},
  doi     = {10.1103/PhysRevD.106.024012}
}

@article{Umarov2025,
  author  = {Umarov, Dilmurod and Atamurotov, Farruh and Ghosh, Sushant G. and Abdujabbarov, Ahmadjon and Mustafa, G.},
  title   = {Dynamics of Spinning Particles around Static Black Holes in Effective Quantum Gravity},
  journal = {European Physical Journal C},
  volume  = {85},
  pages   = {800},
  year    = {2025},
  doi     = {10.1140/epjc/s10052-025-14541-y}
}

@article{Turakhonov2026,
  author  = {Turakhonov, Ziyodulla and Ibrokhimov, Tolibjon and Atamurotov, Farruh and Abdujabbarov, Ahmadjon and Althukair, Ahdab K. and Zotos, Euaggelos E.},
  title   = {Spinning Particles as Probes of Quartic Square-Root Horndeski Gravity: Spin-Curvature Coupling and Orbital Dynamics},
  journal = {European Physical Journal Plus},
  volume  = {141},
  pages   = {629},
  year    = {2026},
  doi     = {10.1140/epjp/s13360-026-07875-3}
}

\end{document}